\documentclass[10pt]{article}
\usepackage[utf8]{inputenc}
\usepackage[margin=1in]{geometry}
\usepackage{graphicx}
\usepackage{color} %
\usepackage{xcolor,colortbl}
\usepackage{setspace} %
\usepackage{newtxmath,newtxtext}
\usepackage{sectsty}
\usepackage{tabu}
\usepackage{longtable}
\usepackage{makecell}
\usepackage{url}
\usepackage{authblk}
\usepackage{lineno}
\usepackage{comment} %
\usepackage{booktabs} %
\usepackage{caption}
\usepackage{defs-private} %



\newcommand{\xhdr}[1]{\vspace{3mm}\noindent{{\bf \large #1}} \\ \noindent}

\newcommand{\todo}[1]{}
\renewcommand{\todo}[1]{{\color{red} TODO: {#1}}}

\usepackage[sorting=none]{biblatex}
\title{Quantifying social organization and\\ political polarization in online platforms}

\author[ ]{Isaac Waller}
\author[ ]{Ashton Anderson}
\affil[ ]{Department of Computer Science, University of Toronto}
\affil[ ]{\texttt {\{walleris,ashton\}@cs.toronto.edu}}
\date{July 2021}

\begin{document}

\maketitle

\begin{refsegment}

{
\bfseries
\noindent Optimism about the Internet’s potential to bring the world together has been tempered by concerns about its role in inflaming the ``culture wars''. 
Via mass selection into like-minded groups, online society may be becoming more fragmented and polarized, particularly with respect to partisan differences~\cite{sunstein2018republic,iyengar2009red}. However, our ability to measure the social makeup of online communities, and in turn understand the social organization of online platforms, is limited by the pseudonymous, unstructured, and large-scale nature of digital discussion. 
We develop a neural embedding methodology to quantify the positioning of online communities along social dimensions by leveraging large-scale patterns of aggregate behaviour.
Applying our methodology to 5.1B Reddit comments made in 10K communities over 14 years, we measure how the macroscale community structure is organized with respect to age, gender, and U.S.\ political partisanship. 
Examining political content, we find Reddit underwent a significant polarization event around the 2016 U.S.\ presidential election, and remained highly polarized for years afterward. 
Contrary to conventional wisdom, however, individual-level polarization is rare; the  system-level shift in 2016 was disproportionately driven by the arrival of new and newly political users. Political polarization on Reddit is unrelated to previous activity on the platform, and is instead temporally aligned with external events. 
We also observe a stark ideological asymmetry, with the sharp increase in 2016 being entirely attributable to changes in right-wing activity. 
Our methodology is broadly applicable to the study of online interaction, and our findings have implications for the design of online platforms, understanding the social contexts of online behaviour, and quantifying the dynamics and mechanisms of online polarization. 
}

\begin{figure}[]
    \centering
    \includegraphics[width=\textwidth]{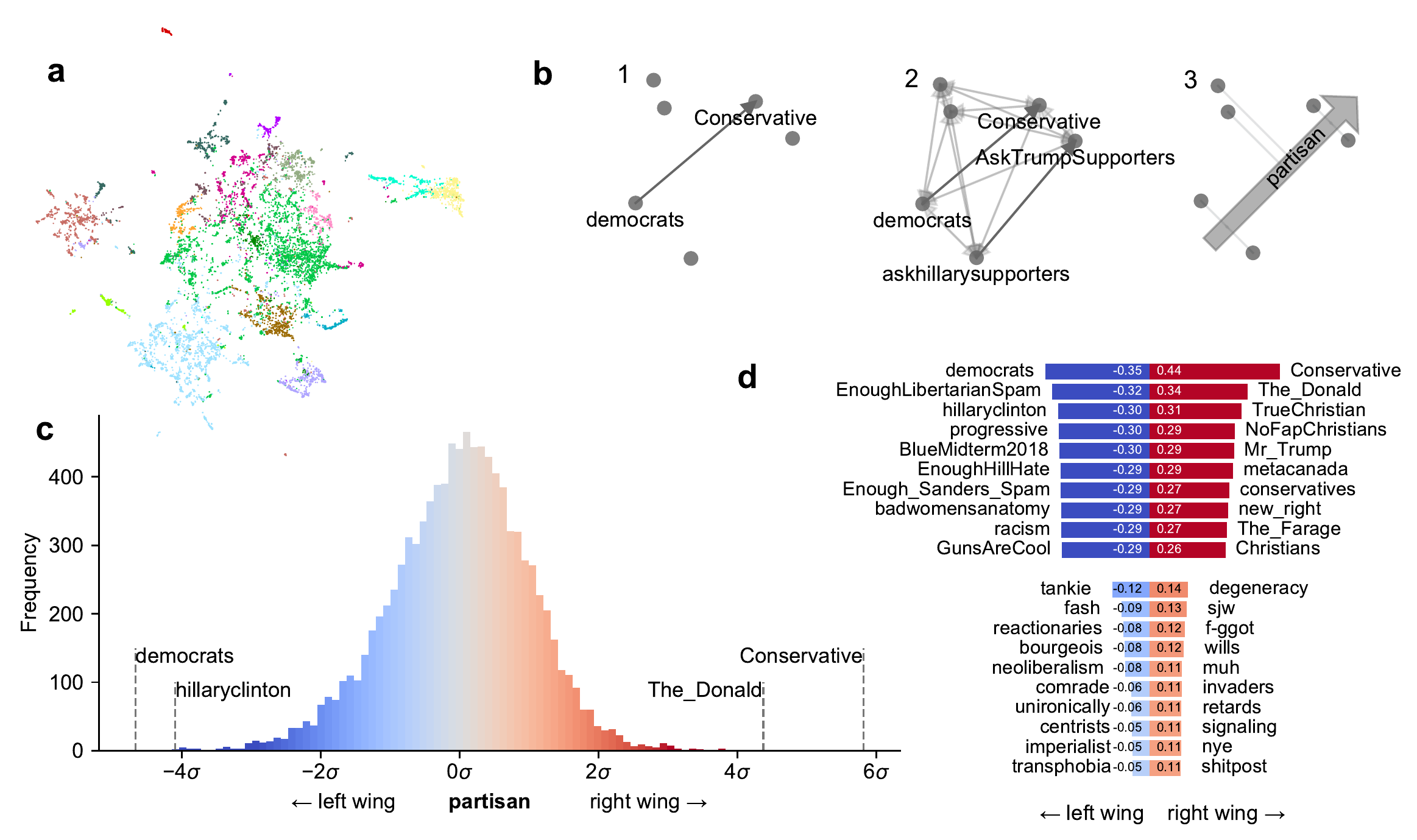}
    \caption{\protect\input{figure-legends/fig1}}
    \label{fig:composite_plot}
\end{figure}

In 1962, Marshall McLuhan proclaimed ``The new electronic interdependence recreates the world in the image of a global village''~\cite{mcluhan1962gutenberg}. In the decades since, there has been fierce debate about the Internet’s dual forces of social integration, as the world becomes increasingly interconnected, and social fragmentation, as people can more easily select into like-minded communities~\cite{van1996electronic,van2013culture,ellison2007benefits}. 
Twenty years into the widespread adoption of online social media platforms, it remains unclear how online communities are socially organized. 
Of particular concern is whether online populations increasingly sort into homogeneous ``echo chambers'', and whether social media platforms tend to shift users towards ideological extremes~\cite{farrell2012consequences,conover2011political,bail2018exposure}. However, since these platforms consist of massive amounts of unstructured and pseudonymous data, empirically quantifying the social makeup of online communities, and in turn the social organization of online platforms, poses a unique challenge.

We develop and validate a methodology using neural community embeddings~\cite{martin2017community2vec}, which represent similarities in community membership as relationships between vectors in a high-dimensional space, to quantify the positioning of online communities along social dimensions. 
Focusing on traditional notions of identity---age, gender, and political orientation---and leveraging the complete set of 5.1B comments made in 10K communities over a 14-year period on Reddit, one of the world’s largest social platforms, we produce an accurate and high-resolution picture of how the platform's macroscale structure is organized along social lines. 
We then apply our methodology to quantify the dynamics and mechanisms of political polarization on Reddit, and investigate three related questions: 
(i) To what extent does platform-level political polarization change over time?, 
(ii) Do individual users become more polarized in their political activity over time, and if so, do these changes drive platform-level polarization?, and 
(iii) Are the dynamics of polarization ideologically symmetric? 

Our approach differs from prior work examining social organization and political polarization in online platforms in three main ways. 
First, our methodology quantifies the social makeup of communities in a purely behavioral fashion. 
Communities are similar to each other if and only if their user bases are surprisingly similar; by computing this similarity along a social dimension (e.g.\ U.S.\  political partisanship), we can recover an accurate estimate of whether a particular community’s user base is more behaviorally aligned with the left or right end of the spectrum (e.g. the left or right wing of U.S.\ politics). 
Users ``vote with their feet'' to decide the social orientation of communities: only action, across large numbers of people, matters. 
In this way, we avoid biases that result from self-reported data~\cite{arnold1981social,van2008faking}, expert labels, and survey-based methods~\cite{chenail2011interviewing,krumpal2013determinants}. 
Prior work has employed word embeddings, high-dimensional representations of text, to study cultural stereotypes~\cite{garg2018word,bolukbasi2016man,caliskan2017semantics} and the cultural markers of class~\cite{kozlowski2019geometry}. Although our dataset is composed of billions of comments, we do not use the text to create our embedding. While differences in identity are reflected in the words people use, this relationship is relatively weak for our focus on measuring the social orientation of underlying community populations. Communities that use similar language may be socially distinct, and communities with distinct language may be socially similar.

Second, previous analyses have studied platforms such as Facebook, Twitter, and Amazon, on which users are guided by algorithmic curation and personalized recommendations~\cite{shi2017millions,del2016echo,conover2011political}. 
As such, traces of user activity on these platforms reflect not only natural human choices but also the influence of underlying algorithms. 
A recent focus has been on examining the effects of algorithmic curation on shaping online social organization, e.g.\ measuring the prevalence of algorithmic ``filter bubbles'' of homogeneous content and groups~\cite{pariser2011filter,flaxman2016filter}, but user choices may play an even larger role in shaping this structure~\cite{bakshy2015exposure}. Thus, although our methodology is generally applicable to many online platforms, we apply it here to Reddit, which has maintained a minimalist approach to personalized algorithmic recommendation throughout its history~\cite{redditcode}. %
By and large, when users discover and join communities, they do so through their own exploration---the content of what they see is not algorithmically adjusted based on their previous behaviour. Since the user experience on Reddit is relatively untouched by algorithmic personalization, the patterns of community memberships we observe are more likely the result of user choices, and thus reflective of the social organization induced by natural online behaviour. 

Finally, we expand the study of political polarization in social media. Polarization is understood as both a state and a process~\cite{dimaggio1996have}, but existing empirical research is largely limited to static analyses of incomplete and non-representative snapshots of platform activity. As such, while the literature has uncovered evidence that online platforms exist in states of partisan fragmentation~\cite{barbera2015tweeting,conover2011political,quattrociocchi2016echo, adamic2005political}, important questions about the dynamics and mechanisms of polarization processes remain unanswered. In particular, the measurement of platform-level polarization with incomplete and non-representative datasets is difficult, and tracking it over time with static analyses is impossible. Furthermore, any observed platform-level polarization could be due to two separate mechanisms with different policy implications: individual users could move towards ideological extremes in their activity over time, or relatively moderate populations could be replaced by new, more extreme populations as the user base turns over. Applying our methodology, we conduct dynamic analyses of complete platform activity to measure both platform- and individual-level polarization, and compare these for the left and right wings, over Reddit’s entire history.

\xhdr{Social dimensions in community embeddings} 
We analyze discussion forum data from Reddit, one of the world's largest online social platforms. 
Reddit is composed of tens of thousands of communities, or ``subreddits'', each of which is typically centered around a single topic or shared interest. 
Subreddits contain posts, which can either be an external link or a short piece of text initiating a discussion, and users can comment on others' posts. 
For our analysis, we use the complete set of comments made on Reddit posts since comments were introduced in 2005 up to and including 2018, collected and distributed by Pushshift~\cite{baumgartner2020pushshift}. 
For all 34.7M Reddit commenters, our dataset contains their complete public commenting history, the communities their comments appeared in, and the timestamps associated with each comment. 

To study the macroscale structure of the platform, we use and extend \emph{community embeddings}~\cite{martin2017community2vec}. 
Much like how word embeddings position words in a high-dimensional space such that similar words are nearby, community embeddings position communities in a high-dimensional space such that similar communities are close together in the space. 
The key difference is that community embeddings are learned solely from interaction data---high similarity between a pair of communities requires not a similarity in language but a similarity in the users who comment in them. 
Communities are then similar if and only if many similar users have the time and interest to comment in them both (a detailed discussion of what similarity entails in this context can be found in the Methods). 
We embed the largest 10,006 communities by number of comments, which account for $95.4\%$ of all Reddit comments, into a 150-dimensional space (\figureSchematic a) and optimize the embedding with community analogies (Methods). 

Analogously to how previous research uncovered axes in word embeddings that correspond to gender, class, and affluence~\cite{  kozlowski2019geometry,bolukbasi2016man,garg2018word}, we develop a methodology to find dimensions in community embeddings that correspond to social constructs. 
To do so, we first identify a seed pair of communities that differ in the target construct, but are similar in other respects. We seed our \dimension{gender} dimension with \subreddit{AskMen} and \subreddit{AskWomen}, question-and-answer forums for men and women; our \dimension{age} dimension with \subreddit{teenagers} and \subreddit{RedditForGrownups}, personal discussion forums for teenagers and adults; and our \dimension{partisan} dimension with \subreddit{democrats} and \subreddit{Conservative}, two partisan American political communities. (Descriptions of every community we reference can be found in Supplementary Table 1.)
To more robustly capture social differences along these dimensions as they are expressed on the platform, we automatically augment these seeds with similar pairs of communities. 
For each dimension, we select the 9 pairs with the most similar vector difference from the set of all pairs of very similar communities (Methods; see Appendix Table 1 for a list of selected pairs.) 
The resulting set of 10 seed vector differences are then averaged together to generate the final dimensions corresponding to each target concept (\figureSchematic b). The method generalizes to more concepts than we study here (Methods).  

Every community can then be positioned along a social dimension  %
by projecting the community's vector representation  %
onto the dimension.  %
This is equal to the focal community's average similarity with communities on the right side of the seed pairs minus its average similarity with communities on the left. 
Communities with memberships that are more similar to one pole end up close to that pole, whereas communities that are equally similar to both ends of the spectrum fall in the middle. 
As the cosine similarity of two communities is related to the similarity between their memberships, a community's score on a dimension is reflective of how similar its membership is with the seeds at either pole. 
The distribution of community scores along the \dimension{partisan} dimension 
varies between the extreme left-wing and extreme right-wing on Reddit (\figureSchematic c). The words most associated with the left and right poles illustrate how political discussion differs across the \dimension{partisan} spectrum (\figureSchematic d).

We validate these dimensions by demonstrating that scores are highly correlated with external measures. 
For the \dimension{gender} dimension, we show that the positions of occupation-based subreddits, such as \subreddit{pharmacy} and \subreddit{Carpentry}, are strongly correlated with the gender proportion of workers in those occupations in a national survey ($r = 0.89$; $p < 10^{-8}$; Appendix Figure 4). 
For the \texttt{partisan} dimension, we analyze communities that explicitly mention their partisan affiliation in their description and verify that their \dimension{partisan} scores reflect their affiliation ($r = 0.92$; Cohen's $d=4.89$; Appendix Figure 5). 
We also verify that the partisan scores of city-based subreddits are correlated with the Republican vote differential in the 2016 U.S.\  presidential election ($r=0.39$; $p < 10^{-5}$; Appendix Figure 4). 
For the \dimension{age} dimension, we observe that communities for universities, e.g.\ \subreddit{UofT}, are consistently far younger than the community for the corresponding city, e.g.\ \subreddit{Toronto} ($r=0.91$, Cohen's $d = 4.37$; Appendix Figure 5). While these validations suggest that the dimensions are correlated with real-world identities, we emphasize that they are measures of social associations, not individual characteristics. A community's position on the gender dimension, for example, should not be interpreted as a direct measure of the gender identity of the community's members, but instead reflects its association with the social constructs of masculinity and femininity as expressed on Reddit (see Methods for more details on validations).

We also generate secondary dimensions that represent the strength of association with each primary dimension, which we term \dimension{partisan-ness}, \dimension{gender-ness}, and \dimension{age-ness}. These secondary dimensions are calculated by taking the sum of the seed pairs' vectors, instead of the difference, and measuring similarity to \textit{both} ends of the primary dimension. For example, \dimension{partisan-ness} corresponds to how political a community is, whereas \dimension{partisan} corresponds to a community's position along the left-right political axis. 
Both \subreddit{progressive}, a community centered on the ``Modern Political and Social Progressive Movement'', and \subreddit{LesbianGamers}, a community for ``women who love women, who love gaming``, are close to the left pole of the \dimension{partisan} axis ($z = -4.0, z = -2.2$), since they tend to have similar memberships as other communities on the left. However, \subreddit{progressive} scores high on the \dimension{partisan-ness} axis ($z = 4.4$) whereas \subreddit{LesbianGamers} scores low ($z = -1.2$). 
We validate the \dimension{partisan-ness} dimension by examining the same explicitly-labeled communities as in the \dimension{partisan} validation, and we find that labeled political communities have far higher \dimension{partisan-ness} scores than communities in general (Cohen's $d=3.27$; Appendix Figure 5).

\begin{figure}[]
    \centering
    \includegraphics[width=\textwidth]{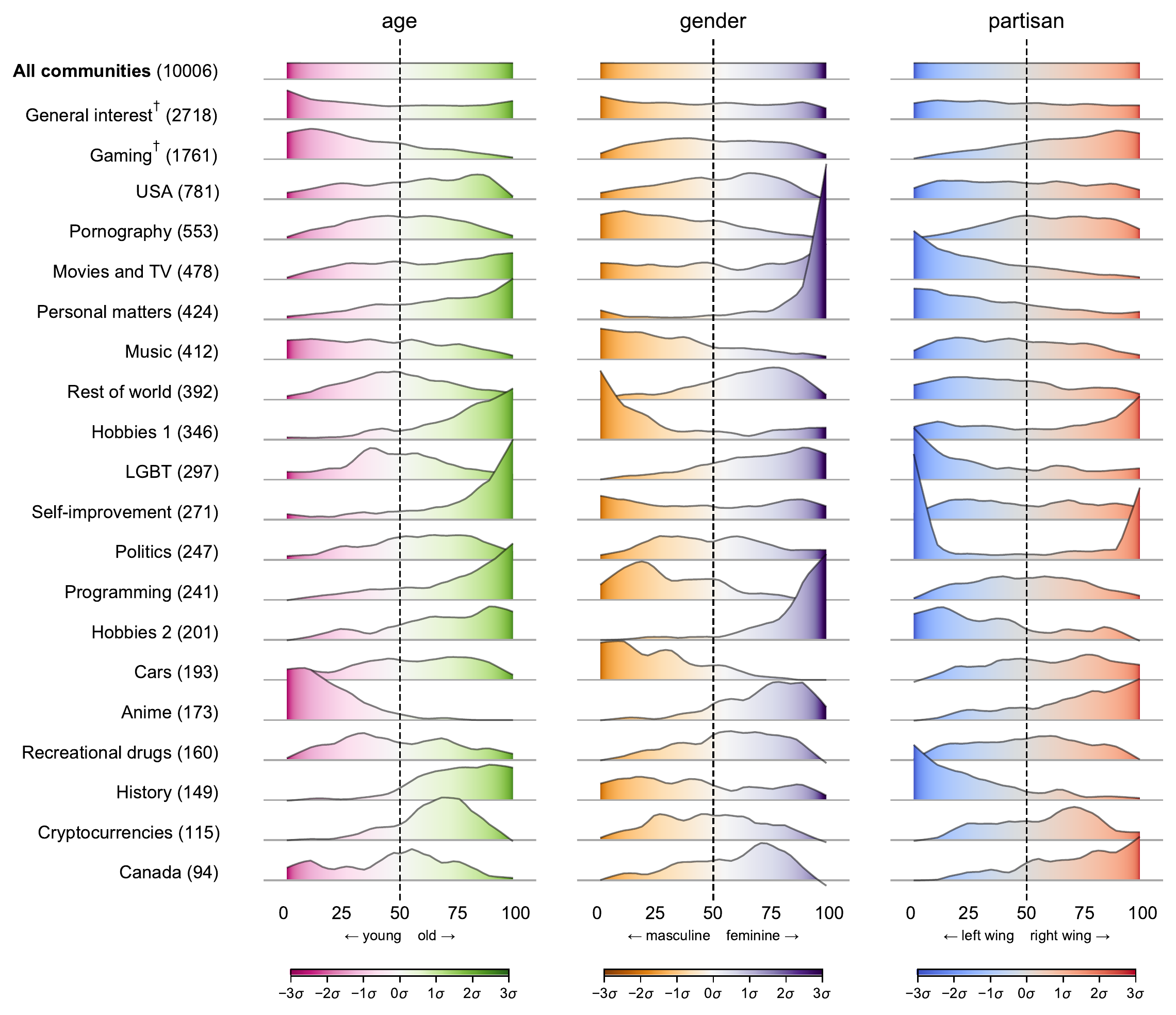}
    \caption{\protect\input{figure-legends/fig2}}
    \label{fig:densities}
\end{figure}

\xhdr{The social organization of Reddit} 
We first apply hierarchical clustering to the embedding to obtain a grouping of communities that reflects the primary similarities and differences in their membership activity, then apply our social dimensions methodology to score every Reddit community along the \dimension{age}, \dimension{gender}, and \dimension{partisan} axes. 
The distributions of Reddit communities along these social dimensions reveal significant inter- and intra-cluster diversity (Figure~2). 
Entire top-level clusters of communities skew strongly towards the poles of the dimensions, significantly departing from the null hypothesis of a uniform distribution over community score percentiles. 
Since the clustering is based on all behavioural relationships in the original community embedding, the top-level clusters could have differed primarily in topic while remaining socially undifferentiated. 
Instead, the stratification along social dimensions demonstrates the importance of age, gender, and U.S.\ partisanship to the high-level organization of activity on Reddit. 
Furthermore, the fine-grained distributions in Figure~2 illuminate how the platform is socially organized. 
For example, Programming communities skew masculine (36\% are below the 20th percentile) and old (50\% are above the 80th percentile),
and Personal matters communities skew feminine (77\% are above the 80th percentile) and left-wing (36\% are below the 20th percentile). 
Hobbies 1 communities skew old (52\% are above the 80th percentile) and masculine (53\% are below the 20th percentile),
while Hobbies 2 communities skew old (39\% are above the 80th percentile)
and feminine (73\% are above the 80th percentile). 
Cars communities skew masculine (45\% are below the 20th percentile) and History communities skew left wing (50\% are below the 20th percentile)
and old (43\% are above the 80th percentile). 
Politics communities exhibit a bimodal distribution on the partisan axis (77\% are below the 20th percentile or above the 80th percentile). 
Additionally, there is substantial diversity within each cluster of communities. Every group has communities that fall on both sides of the global mean of each dimension, and most groups have an outlier community ($>2$ std.\ dev.\ from the mean) on both sides of 0 (\figureManyDensitiesNonpctl). 
The community scores derived from our social dimension methodology offer a high-resolution and large-scale picture of the social makeup of online communities.

To further clarify the nature of Reddit's social organization, we note that the online expressions of social constructs may differ from their traditional meanings in offline contexts. 
Focusing on the \dimension{partisan} axis, we quantify how it relates with the \dimension{gender} and \dimension{age} axes (\figureDimenRelationships a,b) and find that online relationships between constructs do not simply reflect their offline analogues. 
There is a significant monotonic relationship between the \dimension{partisan} and \dimension{gender} dimensions (\figureDimenRelationships a), with communities that skew towards the masculine pole also skewing right-wing ($r=-0.29$). 
The relationship is particularly strong at the \dimension{partisan} extremes; the most left-wing communities are $44.0\%$ feminine-leaning and only $1.4\%$ masculine-leaning, whereas the most right-wing communities are $23.3\%$ masculine-leaning and only $2.9\%$ feminine-leaning. 
At the community level, the political poles on Reddit are almost completely segregated by gender. 
The direction of this relationship is consistent with the American electorate; in the 2016 U.S.\ presidential election, men voted for Trump by a share of 52 to 41, and women voted for Clinton by a share of 54 to 39~\cite{pew2018}. 
We also find a relationship between the \dimension{partisan} and \dimension{age} dimensions (\figureDimenRelationships b);  older communities skew left-wing, while younger communities skew right-wing ($r = -0.37$). 
Among left-wing communities, $38.5\%$ are older but only $2.1\%$ are younger, while among right-wing communities, $26.1\%$ are younger but only $2.9\%$ are older. 
Notably, the direction of this relationship is the opposite of what is traditionally found in offline contexts---in the 2016 U.S.\ presidential election, the 18--29 age group voted for Clinton by a share of 58 to 28, while the 65+ age group voted for Trump 53 to 44---but is consistent with prior observations of the relative youth of the online alt-right movement \cite{hawley2017making}. 
We repeat these analyses on dimensions generated with slightly different seeds to verify the robustness of our method and find similar results (Methods).

\begin{figure}[]
    \centering
    \includegraphics[width=\textwidth]{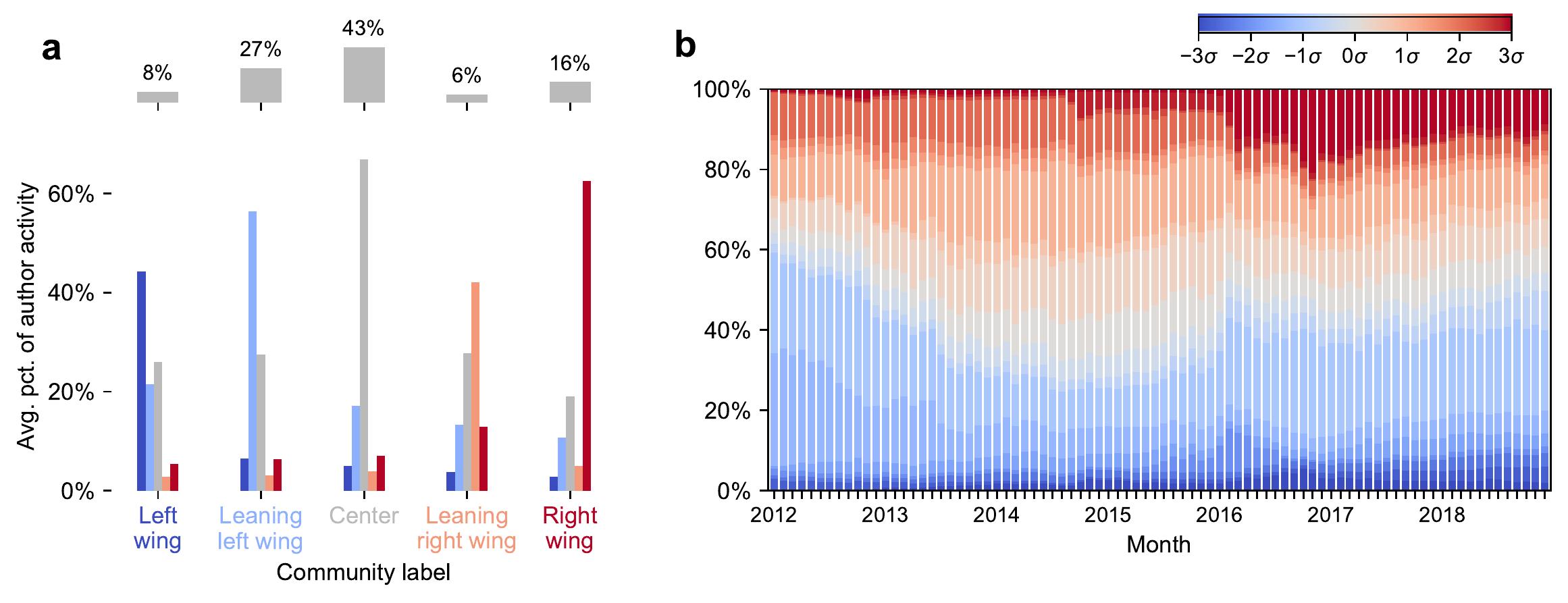}
    \caption{\protect\input{figure-legends/fig3}}
    \label{fig:polarization_1}
\end{figure}

\begin{figure}[]
    \centering
    \includegraphics[width=\textwidth]{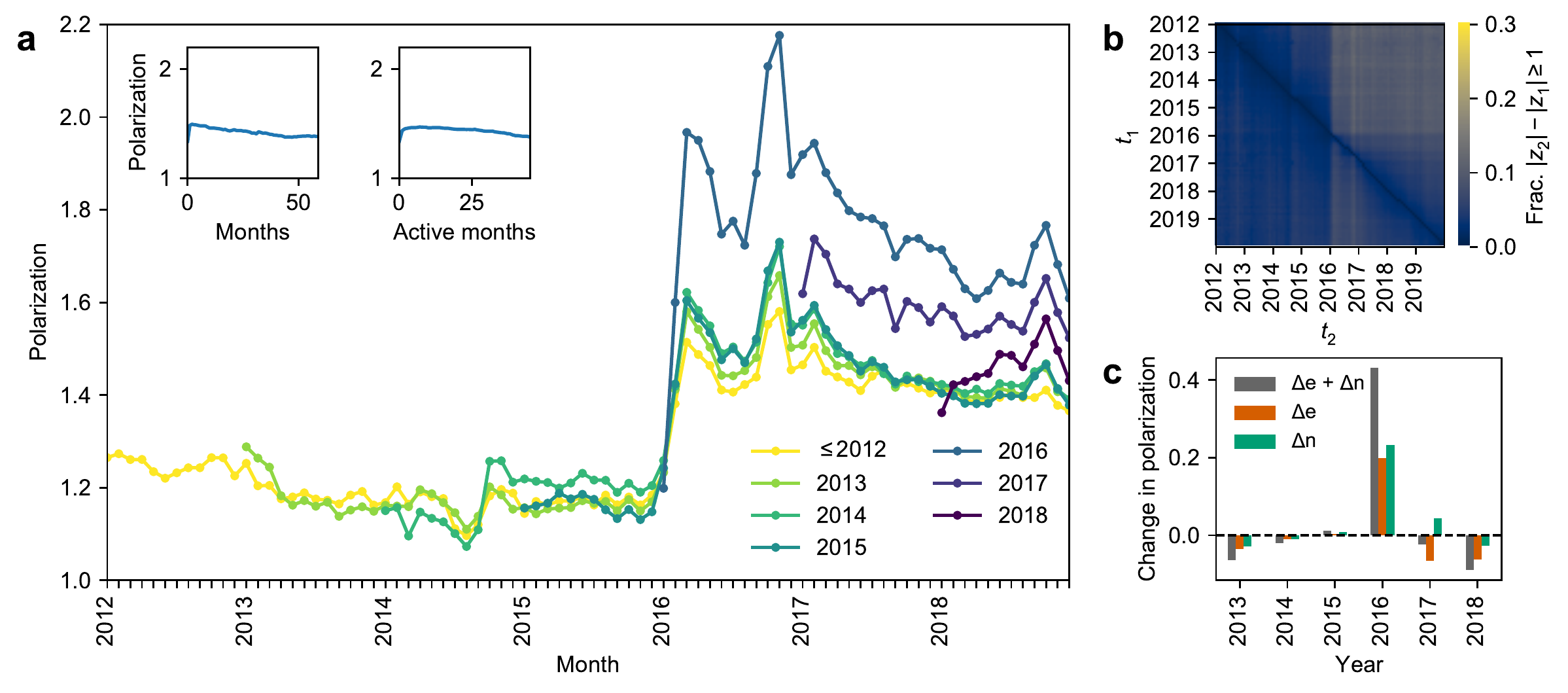}
    \caption{\protect\input{figure-legends/fig4}}
    \label{fig:polarization_2}
\end{figure}
\begin{figure}[]
    \centering
    \includegraphics[width=\textwidth]{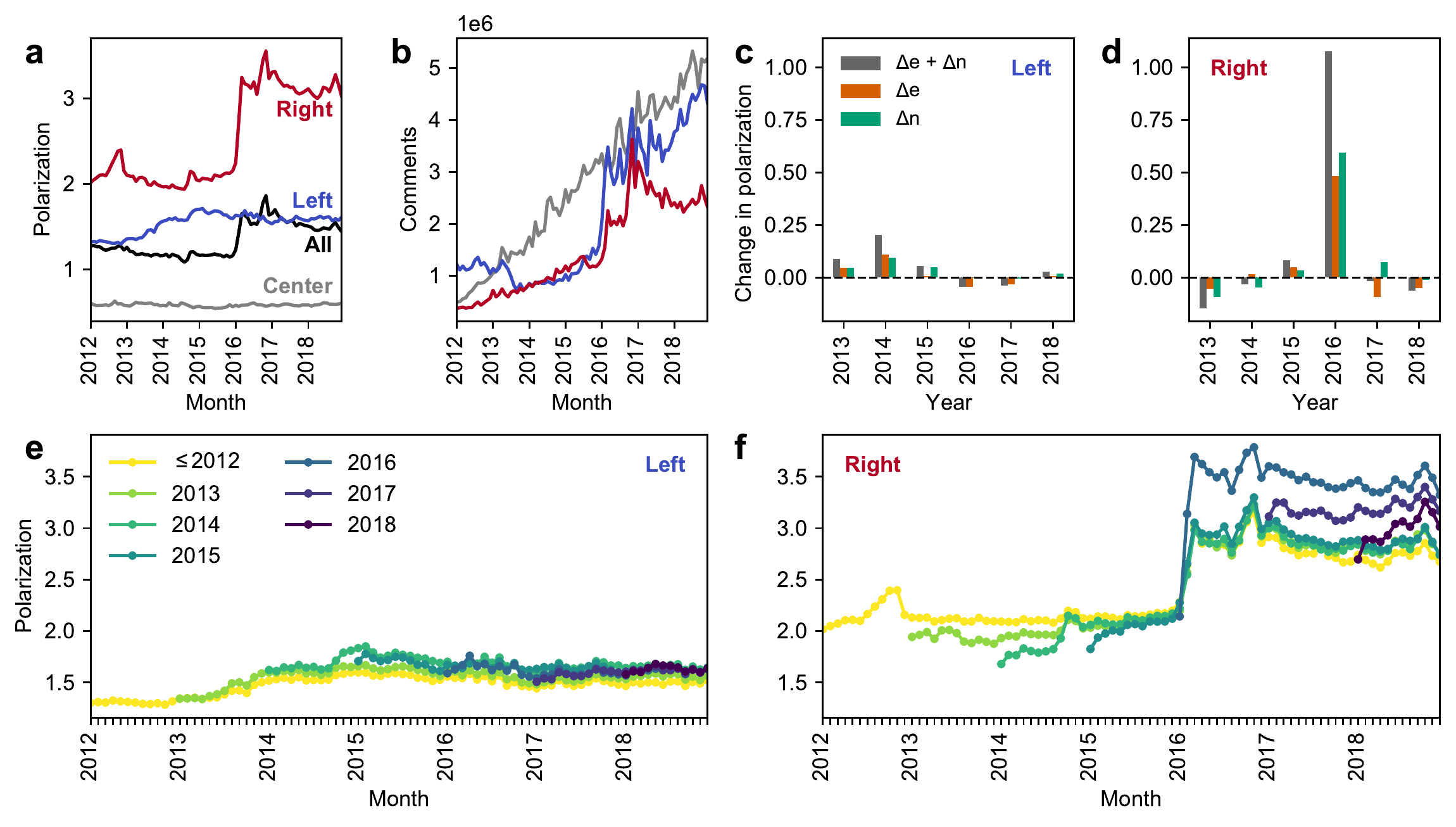}
    \caption{\protect\input{figure-legends/fig5}}
    \label{fig:pol_l_v_r}
\end{figure}

\xhdr{Political polarization on Reddit} 
We now apply our methodology to study individual- and platform-level political polarization over time. 
Political activity on Reddit spans the ideological spectrum, 
with 35\% of activity taking place left of center, 22\% of activity right of center, and 43\% in the center (Figure~3a, top). Despite this overall breadth, user activity is considerably more narrow.
In line with the echo chamber hypothesis, the political activity contributed by a community's members is heavily skewed towards communities with similar partisan scores. 
For example, only 8\% of political discussion occurs in the most left-wing communities, but among users who contribute to the left, an average of 44\% of their activity takes place in left-wing communities. Similarly, only 16\% of political discussion occurs in the most right-wing communities, but the right-wing accounts for on average 62\% of right-wing commenters' political activity. If users' distributions of activity were not skewed along partisan lines, these average percentages would be approximately equal to the overall platform partisan distribution (\figurePolarizationMisc c).
This pattern of selective partisan activity is also clearly apparent at the individual community level (\figurePolarizationMisc d). 
Consistent with previous studies of political activity on social media, a static analysis of complete Reddit activity shows that users selectively participate in ideologically homogeneous communities~\cite{conover2011political,barbera2015tweeting,quattrociocchi2016echo}.

However, this style of analysis cannot address whether selection into partisan communities changes over time. 
To understand whether political activity on Reddit became more polarized throughout the platform's history, we track the distribution of political activity from 2012 to 2018 (Figure~3b). 
While Reddit has always supported a wide range of political activity, the platform became substantially more polarized around the 2016 U.S.\ presidential election. The polarization of discussion, measured by the mean absolute value \dimension{partisan} z-score of political comments (i.e.\ mean absolute number of std.\ dev.\ from the mean), remained consistently within a narrow band between $1.08$ and $1.28$ from 2012 until the end of 2015; it then rose sharply during 2016 and peaked at $1.86$ in November 2016 (\figureLeftvRight a). 
The percentage of political activity that took place in far-left and far-right communities was only $2.8\%$ in January 2015, but peaked at $24.8\%$ in November 2016 (\figurePolarizationMisc a). 
The platform never returned to pre-2016 polarization levels, maintaining monthly mean absolute value \dimension{partisan} z-scores greater than $1.44$ until the end of the data time window.

A central concern is whether individual users become more polarized in their activity over time. %
Overall increases in platform-level polarization could either be driven by individual-level change, with existing users moving towards the partisan extremes, or by population-level turnover, with new users entering the platform in more extreme communities. %
To quantify this, we group users into cohorts based on the date of their first comment in a political community and measure the average polarization of each cohort over time (\figureCohort a). 
This analysis reveals several insights about the dynamics and mechanisms of polarization on Reddit. 
First, with the exception of 2016, users generally do not polarize over time; within-cohort polarization levels usually either remained unchanged or decreased from one year to the next. 
We directly measure individual-level polarization by computing the fraction of users whose activity moved by at least one standard deviation towards the partisan poles. 
This fraction is consistently low; comparing user scores 12 months apart, it was between 1.9 and 3.3\% prior to 2016, and peaked at 11.3\% in November 2016 (\figureCohort b). 
Second, during 2016 every active cohort polarized at the same time. 
The month-to-month polarization trends in 2016 are remarkably synchronized across cohorts. 
Third, the intense increase in polarization in 2016 was disproportionately driven by new and newly political users. 
The change in platform-level polarization was $2.17$ times what it would have been if the 2016 cohort arrived at the average 2015 polarization level, %
despite only accounting for 38\% of political activity during 2016 (\figureCohort c). 
Furthermore, a cohort's increase in polarization was directly related to its age, with newer cohorts polarizing more than older cohorts. 
Finally, individual polarization level is unrelated to previous activity on the platform, either measured by calendar months since first activity (\figureCohort a, left inset) or active months spent on the platform (\figureCohort a, right inset). Changes in polarization over time on Reddit are not associated with previous activity on the platform but rather are synchronously aligned with external events, and are disproportionately driven by new users.

Examining polarization over time separately for left- and right-wing communities reveals a stark ideological asymmetry.  Activity on the right was substantially more polarized than activity on the left in every month between 2012--2018 (Figure~5a). In 2016, discussion on the right shifted significantly rightward, with polarization increasing from an average of 2.12 in November 2015 to a peak of 3.55 in November 2016. %
During the same period, discussion on the left and in the center did not polarize at all (average polarization changed from 1.60 to 1.57 for the left, and from 0.58 to 0.57 for the center). 
The overall platform shift in polarization in 2016 is thus entirely driven by the change in activity on the right, despite the fact that the right is the smallest group by discussion volume (Figure~5b). Similar to the analogous findings for overall polarization, new users on the right in 2016 were significantly more polarized than all prior cohorts and disproportionately drove the observed polarization of the right-wing on Reddit (Figure~5d,f), consistent with the rise of large right-wing communities such as \subreddit{The\_Donald}. Changes in polarization on the left are small by comparison (Figure~5c,e).

Although instances of individual users becoming more polarized in their \dimension{partisan} score over time are rare, it is still possible that newly political users moved from implicit ``gateway communities'' to explicitly partisan communities. 
For example, some communities, such as \subreddit{watchpeopledie}, have a highly partisan user base but are not themselves explicitly political, and thus have extreme \dimension{partisan} scores but low \dimension{partisan-ness} scores (Appendix Figure 6c). 
If engagement with implicitly left- or right-wing communities is related to an increased propensity to subsequently engage with explicitly left- or right-wing communities, this could be evidence of an implicit process of polarization occurring on the platform. 
However, for users who were active in an explicitly left-wing or right-wing community, in any given month at most $27\%$ had contributed in a previous month to an implicitly left-wing community and $27\%$ to an implicitly right-wing community, restricting the population for whom such an effect could apply (Appendix Figure 9 top). 
Users tend to become active in both implicitly and explicitly partisan communities in the same month, further indicating that such a polarization effect is limited in its possible impact (Appendix Figure 9 bottom).

There are limitations in our methodological approach. For example, by representing each community by a single vector in a common embedding, we measure community relationships aggregated over the entire time period of our dataset. This implicitly assumes that community similarities, and community scores on social dimensions, do not change. 
Although it is plausible that some communities change significantly in the social makeup of their membership, we expect these to be exceptional cases as most communities are organized around a fixed topic or shared interest. 
Our method relies on co-membership data---examples of the same user being a member of several communities. 
If large numbers of people use ``throwaway'' user accounts for certain communities (e.g.\ those dedicated to fringe or controversial content), thereby splitting their activity over several accounts, the relationships between these communities and the rest of the platform could be distorted.

Group membership is fundamental to social identity.  Sociologists dating back to Simmel, who pioneered the notion of ``the web of group affiliations'', have employed complex characterizations to understand identity~\cite{simmel2010conflict,breiger1974duality,bourdieu1984distinction,feld1981focused,crenshaw2017intersectionality}. 
We have shown that by harnessing mass co-membership data, we can use high-dimensional representations of online communities to quantify behavioural similarities and differences in their memberships along key social dimensions. 
Applying our methodology to a complete longitudinal dataset of political activity on Reddit, we find the platform underwent a significant polarization event around the 2016 U.S.\ presidential election. 
However, individual-level polarization is rare, and platform-level polarization is disproportionately driven by new and newly political users. 
Polarization on Reddit is unrelated to previous activity on the platform, and is instead synchronized with external events. 
We also find a clear ideological asymmetry, with the 2016 spike in polarization being entirely attributable to increasingly extreme activity on the right. 
These findings shed light on the dynamics and mechanisms of political polarization in social media. 

This study introduces a new paradigm for the analysis of online platforms. 
Embedding communities into a high-dimensional space and projecting them onto dimensions that correspond to social constructs distills vast amounts of behavioural metadata into semantically meaningful, fine-grained, and valid measurements of social alignment.  
Our methodology can be generally applied to quantify the social organization of online discussion, to situate important content and behaviours, such as misinformation and toxic language, in the social context of online platforms, and to quantify the nature of individual- and platform-level online polarization and the mechanisms that drive it.

\nolinenumbers

\printbibliography[segment=1,heading=none]
\end{refsegment}

\section*{Methods}

\begin{refsegment}

\xhdr{Data} For our analysis, we use the complete set of 5.1B comments made on Reddit posts since comments were introduced in 2005 up to and including 2018. The dataset is publicly available and was downloaded from the \texttt{pushshift.io} Reddit archive \cite{baumgartner2020pushshift} at \url{http://files.pushshift.io/reddit/}. For all 34.7M Reddit commenters, our dataset contains their complete public commenting history, the communities their comments appeared in, and the timestamps associated with each comment. Over our entire study period, $52.9\%$ of users commented in more than one subreddit, and the mean number of subreddits commented in by a user is $9.6$, demonstrating that many users engage in the multi-community aspect of the platform. This activity provides crucial information about the behavioral similarity of subreddits, which we harness to create community embeddings and social dimensions.

\xhdr{Creating the community embedding} We use this Reddit commenting data to represent communities in a behavioural space using community embeddings, which were first proposed by Martin~\cite{martin2017community2vec} and were subsequently refined by Kumar et al.~\cite{kumar2018community} and Waller and Anderson~\cite{waller2019generalists}. We create a community embedding from the Reddit data set using the open source software \texttt{word2vecf} (\url{https://bitbucket.org/yoavgo/word2vecf/src}), a modification of the original \texttt{word2vec} software to allow the usage of arbitrary contexts~\cite{levy2014dependency}. To generate our embedding, we apply the word2vec algorithm to interaction data by treating communities as "words" and users as "contexts"---every instance of a user commenting in a community becomes a word-context pair. %
For example, if user $u_i$ commented in community $c_j$ 10 times, the pair $(u_i, c_j)$ would appear 10 times in the training data. The model is then trained using the skip-gram with negative sampling (SGNS) method. To remove extremely small subreddits for which there is insufficient data to generate a meaningful vector representation, we restrict to the top 10,006 subreddits by number of comments, which accounts for $95.4\%$ of all comments and $93.2\%$ of all users. Since our training data is generated without using a context window or intermediate documents, in contrast with traditional \texttt{word2vec}, all word-context (user-community) pairs are included in our training data without restriction (analogous to using an infinite-size context window).

The \texttt{word2vecf} model has numerous hyperparameters that affect the training process and resulting embedding. To tune the model for the community embedding use case, we perform a grid search of the hyperparameter space, optimizing for performance on a set of community analogies. The hyperparameters we vary are: \texttt{sample}, the down-sampling threshold; \texttt{negative}, the number of negative examples; \texttt{alpha}, the starting learning rate; and \texttt{size}, the dimensionality of the resulting embedding. We add an additional parameter \texttt{shuffled}, a Boolean parameter which indicates whether the training data should be randomly shuffled prior to training. We assess the model's performance on three sets of analogies: university subreddits to their corresponding cities; sports teams to their corresponding cities; and sports teams to their corresponding sport. 
By performing a grid search of the hyperparameter space, we find an embedding that solves 72\% of the 4,392 analogies perfectly, and 96\% of them nearly perfectly (correct answer in the top 5 communities.) The resulting parameters from this process are \texttt{-alpha 0.18, -negative 35, sample 0.0043, -size 150, -shuffled true}. We believe that shuffling the data set prior to training prevents the model from over-fitting on temporal trends.

SGNS learns not only a vector for each word (in this case, each community) but a vector for each context as well (in this case, each user). While we only use the word (community) vectors in this paper, the context vectors play an important role in the training process. The training objective of the SGNS training procedure maximizes the dot product of word-context pairs that frequently co-occur, and minimize the dot product of randomly generated word-context pairs (negative examples). Intuitively, this suggests that communities with `similar' users will end up with similar vectors, and users who participate in `similar' communities will end up with similar vectors. However, this circular definition does not provide a concrete interpretation for the dot product of two community vectors. Levy and Goldberg \cite{levy2014neural} show that the SGNS objective is optimized by a factorization of the word-context pointwise mutual information (PMI) matrix (shifted by a constant). PMI is a measure of association between a word and a context, or, in our context, a measure of association between a community $c$ and a user $u$, where $\#$ is the count of all matching comments:

\[
    PMI(c, u) = \log \frac{P(c, u)}{P(c)P(u)} = \log \frac{\#(c, u) \cdot \#_{total}}{\#(c) \cdot \#(u)}
\]

Note that this matrix is dense, and in the common case where $\#(c,u)=0$, $PMI(c,u)=-\infty$. In such a PMI matrix, the dot product of two community vectors is related to the similarity of their PMI values over all users:

\[
    \vec{c_1} \cdot \vec{c_2} = \sum_{u} PMI(c_1,u) \cdot PMI(c_2, u)
\]

If SGNS was truly a pure factorization of the word-context PMI matrix, it would follow that this approximately holds in a community embedding as well. However, the iterative nature of the training procedure means that SGNS captures not only literal user overlap between communities but higher-order similarities as well. For example, if the two communities \subreddit{trucks} and \subreddit{golf} had no users in common, but both had a high overlap with the \subreddit{AskMen} community, their vectors might end up somewhat close to each other despite no users being members of both communities. Indeed, empirical tests of SGNS and PMI demonstrate that SGNS is extremely capable of preserving second-order context overlap---even weighting this higher than first-order context overlap---while PMI is completely incapable of capturing it at all. In a simulation experiment performed by Schlechtweg et al.~\cite{schlechtweg2019second}, the average cosine distance between words with first- and second- order context overlap were 0.11 and 0.00 respectively using SGNS and 0.51 and 1.0 using PMI.
While matrix factorization of the PMI matrix is also able to capture such higher-order effects, Levy and Goldberg establish that in practice SGNS arrives at a different result than factorization of the PMI matrix, and that pure factorization does not perform well on many NLP tasks \cite{levy2014neural}.
Thus, while deriving a closed-form equation that relates the cosine similarity of communities to their actual user overlap is still an unsolved problem, the architecture of the training process and empirical evidence suggests that cosine similarity of two community vectors is a strong measure of the similarity of the user-bases of the two communities.

We perform a clustering of the community embedding to understand Reddit's macroscale community structure. We use agglomerative clustering based on Euclidean distance to partition all communities into 30 clusters. We then manually label the clusters based on their dominant topic, e.g.\ Movies and TV ($n=478$), Music ($n=412$), and Politics ($n=247$). 
When more than one cluster has the same topical theme, we label them in descending order of size, e.g.\ Hobbies 1 ($n=346$) and Hobbies 2 ($n=201$). 
Six clusters consist of communities with no clear theme, which we label General interest (1 through 6).
To conserve space in \figureManyDensities, we merge the six General interest clusters into a single General interest row and the five Gaming clusters into a single Gaming row.

\xhdr{Finding social dimensions} Our methodological contribution is the idea and technique of finding social dimensions in community embeddings that correspond to social constructs. These dimensions allow us to compute scores that represent the social makeup of online communities. We first describe the generic algorithm for constructing social dimensions, then discuss the particular choices we made in our analyses. In the following sub-sections, we describe the computation of community scores and validate them against both internal and external sources.

To generate a social dimension that corresponds to a social construct, the analyst first identifies a seed pair of communities that differ primarily in the target construct. An ideal choice of seed is a pair of communities that are extremely similar except for a difference in the target social dimension. Note that the seed pair communities do not need to be at the extreme ends of the target dimension; they only need to differ primarily in the social construct. 

Second, to ensure that the dimension is not overly tied to idiosyncrasies of the two seed communities, the seed pair $(s_1, s_2)$ is algorithmically augmented with additional similar pairs of communities. Let $k$ denote the desired total number of pairs, chosen by the analyst. We generate the set of all pairs of communities $(c_1, c_2)$ such that $c_1 \neq c_2$ and $c_2$ is one of the 10 nearest neighbours to $c_1$. This is based on the aforementioned idea that we are looking for pairs of communities that are very similar, but differ only in the target concept. All pairs are ranked based on the cosine similarity of their vector difference with the vector difference of the seed pair $\cos (\vec{s_2}-\vec{s_1}, \vec{c_2}-\vec{c_1})$. 
Additional pairs are then selected greedily. 
The most similar pair to the original seed pair that has no overlap in communities with the seed pair or any of the previously selected pairs is selected, and this process is repeated until 
$k-1$ additional pairs are selected, which results in the $k$ pairs used to create the dimension.

Third, the vector differences of all $k$ pairs are averaged together to obtain a single vector that robustly represents the desired social dimension. We also compute a complementary \dimension{-ness} version of the dimension by averaging the vector \textit{sums} of all $k$ pairs. This dimension represents similarity to the communities on both sides of the pairs. 

In our analysis, we choose $k=10$, which implies that $k-1=9$ additional pairs are chosen to augment each seed pair. 
We tested with more and fewer than 10 pairs; fewer and axes appeared to be less robust, and more produced extremely similar axes (by cosine similarity and correlation between scores.) Using fewer pairs allows for conclusions to be drawn about more communities, so we opted for the fewest pairs with good robustness.
We generate the set of all 100,060 non-trivial pairs of communities $(c_1, c_2)$ with their 10 nearest neighbours and build dimensions as described above. 
For our \dimension{gender} dimension, we choose \subreddit{AskMen} and \subreddit{AskWomen}, personal discussion forums for men and women; for our \dimension{age} dimension, we choose \subreddit{teenagers} and \subreddit{RedditForGrownups}, personal discussion forums for teenagers and adults; and for our \dimension{partisan} dimension, we choose \subreddit{democrats} and \subreddit{Conservative}, two partisan American political communities. 
While we focus here on traditional forms of identity, the method is not inherently constrained to one-dimensional representations. For example, multiple gender dimensions could be generated to build a more complete analysis of gender. Appendix Table 1 contains the 9 similar pairs automatically found for all the dimensions.

While the choice of seed is important, our dimension generation method is robust, as similar seed choices generate similar dimensions. To demonstrate this, we also generate a \dimension{gender B} dimension with \subreddit{Daddit} and \subreddit{Mommit}, parenting discussion forums for men and women; an \dimension{age B} dimension with \subreddit{AskMen} and \subreddit{AskMenOver30}, Q\&A communities for men of all ages and men over 30; and a \dimension{partisan B} dimension with \subreddit{hillaryclinton} and \subreddit{The\_Donald}, two partisan American political communities. 

As an additional notion of identity, we generate an \dimension{affluence} dimension, choosing as seeds \subreddit{vagabond}, a forum for homeless travellers, and \subreddit{backpacking}, a more general interest travel community. We also generate three dimensions for concepts not necessarily related to traditional identity but relevant to Reddit as a platform: \dimension{time}, representing actual time from 2005 to the present; \dimension{sociality}, representing how discussion- and meetup-focused a community is; and \dimension{edginess}, representing provocation and antagonism (seeds can be found in Appendix Table 1).

\xhdr{Computing community scores} Once a vector for a dimension has been obtained, all communities can be assigned a score on that dimension by simply projecting the normalized community vector $\vec{c}$ onto that vector: $\vec{c} \cdot \vec{d}$. The score of a community on a dimension is proportional to its average similarity with the right side minus its average similarity with the left side. This can be seen by noticing that the cosine similarity of a normalized community vector $\vec{c}$ with a social dimension with $n$ normalized seed pairs $(A_1, B_1) ... (A_n, B_n)$ defined as $\vec{d} = \frac{1}{n}\sum(B_i - A_i)$ is the following:

\[
    \cos(\vec{c}, \vec{d}) = \frac{\vec{c} \cdot \sum(B_i - A_i)}{n \lVert \vec{d} \rVert} = \frac{1}{n \lVert \vec{d} \rVert}\sum(\vec{c} \cdot B_i - \vec{c} \cdot A_i)
\]

As previously explained, the dot product of two community vectors is a measure of the similarity of their members. Thus, a community much more similar to one seed than the other will have a score at the poles, while a community equidistant between each of the seeds would receive a score of $0$.
 
We calculate the scores for all 10,006 communities on all dimensions. The distributions of community scores for \dimension{age}, \dimension{gender}, \dimension{partisan}, and \dimension{affluence} can be found in \figureAllDists . 
Distributions broken down by semantic cluster for \dimension{age}, \dimension{gender}, and \dimension{partisan} can be found in \figureManyDensitiesNonpctl~and for %
\dimension{affluence}, \dimension{time}, \dimension{sociality}, and \dimension{edginess} in \figureManyDensitiesExtra .

To demonstrate the robustness of the dimension generation method, we compare each of the \dimension{age}, \dimension{gender}, \dimension{partisan} axes with their \dimension{B} version. 
The \dimension{age} dimension is correlated with \dimension{age B} at $r = 0.90$; \dimension{gender} is correlated with \dimension{gender B} at $r=0.86$; and \dimension{partisan} is correlated with \dimension{partisan B} at $r=0.55$. These results demonstrate that community scores are robust to small changes in the input seeds. The \dimension{partisan B} dimension has a more moderate correlation than the other two. This is because \dimension{partisan} and \dimension{partisan B} capture slightly different concepts. 
For example, Trump was an outsider candidate and online Trump supporters displayed significantly different behaviour than the traditional online Republican base. Therefore, using \subreddit{The\_Donald} as a seed generates a dimension that is more specific to Trump and his online supporters' interests, in contrast with using \subreddit{Conservative} as a seed, which generates a dimension that more closely captures Republicanism in general. This emphasizes the importance of validating community scores using external constructs, as we do in the next section.

\xhdr{Validating community scores} We validate each of \dimension{age}, \dimension{gender}, \dimension{partisan}, and \dimension{partisan-ness} against the external concepts they represent. To validate the \dimension{gender} dimension, we compare the \dimension{gender} scores of occupation communities to the actual gender makeup of those occupations. We use gender makeup data from the 2018 American Community Survey, and manually match occupation descriptions to subreddit names (\tableGenderValidation .) We find there is a $r=0.89$ correlation between the percentage of women in an occupation and its communities' gender score (see scatter plot in \figureValidationsScatter .) The \dimension{gender} dimension well represents the proportion of women in an occupation even for occupations at the extremes and in the middle. To validate the \dimension{age} dimension, we compare communities for universities and the communities for the respective cities, as universities tend to have a much younger population than a city as a whole. We find a very strong relationship between \dimension{age} and whether a community is associated with a university or a city ($r = 0.91$, Cohen's $d=4.37$). As shown in \figureValidationsAgePartisan , university communities skew far younger and city communities skew far older.

To validate the \dimension{partisan} dimension, we manually code communities as left or right wing, and verify that the partisan score distinguishes between them. We select communities that contain in their description either one of the left-wing terms ``democrat'', ``clinton'', ``left'',  ``progressive'' or one of the right-wing terms ``republican'', ``trump'', ``right'', ``conservative''. We then manually code these communities based on their description into one of two categories: left-wing (or anti-right) and right-wing (or anti-left). Coding is performed strictly using these words and whether the description is supportive or against them. We code 125 communities which contain one of these words and find 32 left-wing and 18 right-wing communities. The remainder were not labelled as there was no clear association in the description. We find that this label is strongly associated with the partisan score ($r = 0.92$, Cohen's $d=4.89$). We also use this labelling to validate the \dimension{partisan-ness} dimension. We compare the distribution of \dimension{partisan-ness} scores for the labelled left or right communities and find it is substantially different than that of all other communities (Cohen's $d=3.27$).

We perform an additional validation using 2016 US Census data for the \dimension{affluence} and \dimension{partisan} dimensions. Reddit communities are matched to US Census metropolitan statistical areas (MSAs) by manual coding. We find that the median household income in a MSA is associated with the \dimension{affluence} score of MSA communities ($r=0.39$), and the Republican-Democrat vote differential in the 2016 presidential election (calculated for each MSA by combining county-level results from the MIT Election Lab) is associated with the \dimension{partisan} score of MSA communities ($r=0.42$). 
The presence of this correlation indicates that our method captures online partisanship, although see \figureDimenRelationships~and the related discussion in the main text for more on how the online expression of U.S.\ partisanship differs from its traditional offline analogue, including voting patterns in presidential elections.

\xhdr{Measuring relationships between dimensions} After validating the social scores, we measure the relationships between these dimensions as they exist on Reddit. %
We find a weak correlation exists between \dimension{age} and \dimension{gender} ($r=0.10$); a moderate correlation exists between \dimension{gender} and \dimension{partisan} ($r=-0.29$); and a moderate correlation exists between \dimension{age} and \dimension{partisan} ($r=-0.37$). 
We repeat this analysis on the alternate B axes for robustness. We find similar relationships between partisan B and gender ($r=-0.34$), between partisan B and age ($r=-0.13$), between partisan and gender B ($r=-0.26$), and between partisan and age B ($r=-0.33$).
\figureDimenRelationships~illustrates the relationships between \dimension{partisan} and \dimension{age}, \dimension{partisan} and \dimension{gender}, and  \dimension{partisan} and \dimension{partisan-ness}.

\xhdr{Computing word scores} We additionally compute scores for words along all dimensions to provide context to our primary analyses. Word scores are weighted averages of community scores weighted by the number of times the word was used in a community. To avoid distortion introduced by bots that re-use the same word over and over again in automated postings, we cap the number of usages of a word in a subreddit by one commenter that are counted at 100. Word scores represent the types of communities in which that word is likely to be observed. The words with the most extreme scores on each of our primary axes are available in Appendix Figure 1.

\xhdr{Measuring political polarization} To quantify political polarization on Reddit,%
we first restrict our focus only to "explicitly political activity"---comments in political communities as defined by the \dimension{partisan-ness} axis. We choose a cutoff on the \dimension{partisan-ness} axis such that it is the highest value that includes 80\% of the "Politics" cluster. Using this cutoff to categorize communities as explicitly political, we label 553 (5.53\%) of communities as political, and we find that it correctly categorizes 92\% of the communities manually coded as  `political' by us based on their description in the previously described validation for the \dimension{partisan} dimension. For each political community, we calculate its partisan z-score $z$ from its partisan score $\vec{c} \cdot \vec{d}$ and the mean and standard deviation of the entire \dimension{partisan} distribution (including non-political communities): $z = \frac{(\vec{c} \cdot \vec{d})-\mu}{\sigma}$. $z$ represents the partisan association of a community, with a $z$ of $0$ indicating that a community has a \dimension{partisan} score equal to the overall mean (i.e.\ it is in the center), negative scores indicating a left-wing association, and positive scores indicating a right-wing association. The \dimension{partisan} z-score of a comment is equal to the \dimension{partisan} z-score of the community it was posted in.

We further restrict our attention to the 88.8\% of political comments which have not been deleted. Deleted comments on Reddit are still visible, but their author is hidden. As we are lacking author data for these comments, we are unable to tell whether they were made by a new user or an existing user. Since one of our aims is to attribute changes in activity based on the prior political activity of users, we exclude these deleted comments from our political analyses. While deleted comments account for a small fraction of overall political activity, it is possible that deleted comments differ from non-deleted comments to such an extent that it affects our main findings. To assess whether such a difference exists, we compare the distribution of partisan scores of deleted comments $Q$ to the distribution of partisan scores of non-deleted comments $P$. The distributions are extremely similar (\figurePolarizationMisc a); they have a difference of means of only $-0.01$ and a Kullback–Leibler divergence of $D_{KL}(P \parallel Q)=0.033$ bits. We conclude that it is reasonable to exclude deleted comments from our analyses.

To measure the extent to which users self-select into partisan groups, we assign all political communities one of five bins $B \in \{-2,-1,0,1,2\}$ by z-score on the partisan axis (left wing (-2): $z<-2$, leaning left (-1): $-2<z<-1$, center (0): $-1<z<1$, leaning right (1): $1<z<2$, right wing (2): $z>2$). 
The proportion of all political activity that falls in each of these bins yields a discrete distribution of political activity on Reddit (\figurePoliticalActivity a top).
Within each bin $b_1$, we measure the likelihood that, if one randomly draws a comment in bin $b_1$, and then randomly draws one of its author's comments, the latter drawn comment falls in bin $b_2$. This measure is designed to give an idea of how much users that contribute to one bin contribute to the same or other bins and is equivalent to the average proportion of activity by authors in $b_1$ that takes place in the bin $b_2$. When $b_1 = b_2$, this can be interpreted as the average proportion of activity by authors in $b_1$ that takes place in the same bin.
Let $A$ denote the set of all authors. Let $c_{a,b}$ denote the number of comments made by author $a$ in bin $b$. The average proportion of activity that takes place in bin $b_2$ by authors in $b_1$ is therefore:
\[
    f(b_1, b_2)=\frac{1}{\sum_{a\in A} c_{a,b_1}}\sum_{a\in A}c_{a,b_1}\frac{c_{a, b_2}}{\sum_{b \in B} c_{a,b}}
\]

Notice that this quantity is weighted by the number of comments an author makes in a bin. Were authors not weighted by their number of comments, authors that make many comments in one bin and a non-zero but small amount of comments in other bins would influence all distributions equally, making all distributions look artificially similar.
We also compute this on the community level, where an individual community is substituted in the place of $b_1$. Results of the community-level analysis are shown in \figurePolarizationMisc c.

If each users' individual distribution over partisan was equal to the overall distribution, i.e.\ there was no self-selection into partisan groupings, each bin's distribution would be approximately equal to the overall activity distribution (\figurePoliticalActivity a top). In such a scenario, where all users had the same likelihood to contribute to a bin, we would still expect to observe slightly more average activity in the 'same bin' in the above analysis due to two factors: one, in order to be included in the calculation for a bin a user must have contributed to it and therefore that users with no activity in a bin are excluded from its calculation, and two, since we choose the bins based on score on the partisan axis, communities within a bin are more similar to each other than average communities, and similarity in the embedding is correlated with user overlap. To show these effects are negligible, we repeat this analysis on a random dataset, generated by randomly shuffling all of the authors of Reddit comments.
Since the userbases of all communities are similar in this random dataset, community vectors tend to be similar in the resulting embedding. As a result, there is far less variation in partisan score among political communities in this embedding, making it impossible to use the previous method of labelling communities by partisan affiliation by standard deviations from the mean. We instead create a best approximation to the conditions in the real embedding by selecting the same number of political communities (i.e.\ we take the 553 communities with the highest \dimension{partisan-ness} scores as `political') and then dividing these communities into five bins of the same number of communities as in the actual embedding by choosing the appropriate thresholds on the \dimension{partisan} axis in the random embedding. This accomplishes the goal of selecting bins with similar \dimension{partisan} scores to put an upper bound on the possible effects of the aforementioned confounds.
The results in this random dataset show that all bin distributions are extremely similar to the overall distribution with a small (less than 0.85\%) increase in the average percent for the same bin, showing that the overall activity distribution is an accurate reference point for what bin distributions would look like were there no self-selection into partisan groupings.

\xhdr{Measuring dynamics of polarization} To measure how platform-level polarization has changed over time, we measure the distribution of political activity on the partisan axis over time.
Again focusing on only the subset of non-deleted comments in explicitly political communities, we quantify the distribution of \dimension{partisan} scores each month. \figurePoliticalActivity b displays the distribution of the \dimension{partisan} scores of comments each month.
As a direct measure of the partisan polarization of the distribution, we also compute the average absolute \dimension{partisan} z-score $|z|$ of activity in each month, i.e.\ the average number of standard deviations from the mean \dimension{partisan} score, for each month (\figureLeftvRight a). Note that we use the average absolute z-score and not the absolute average z-score. Using the absolute average z-score, equal amounts of activity in the far left and far right would average out to zero and be considered non-polarized. As we wish to capture the extent to which activity takes place in polarized communities irregardless of polarity, we use the average absolute z-score.
As an alternate metric, \figurePolarizationMisc b displays the proportion of activity that takes place in very left- and right-wing communities in any given month; very left-wing communities are those with a z-score less than $-3$ (42 communities), while very right-wing communities are those with a z-score greater than $3$ (24 communities).

To measure the extent to which individuals have moved towards partisan extremes as they act on the platform, and the extent to which this has contributed to the overall platform polarization observed, we analyze the distribution of political activity of users with different levels of past activity. We divide all Reddit users active in political communities in ten cohorts by the year they made their first comment in political communities. %
To measure the average polarization of a cohort's activity, we use the average absolute \dimension{partisan} z-score $|z|$.  \figureCohort a illustrates the average absolute z-score of each cohort's activity over time. As an alternate way to visualize the relationship between users' past and present activity, we plot a version of \figurePoliticalActivity b broken down by users' prior political activity in \figurePoliticalActivityOriginal.

We compute two alternate measures of a user's time on the platform to provide a comparison point for the above analysis. For each comment in a political community, we compute the number of calendar months since the author's account was created, $a$, and the number of distinct calendar months the author has been active in political communities up to the point of the comment's posting, $b$. We group political activity by $a$ and $b$ and calculate the average absolute z-score for these comments. \figureCohort a insets display the relationship between $a$ (left) or $b$ (right) and the average absolute z-score of political comments.

To determine how common it is for users to significantly polarize in activity, we compare the same user's political activity in different calendar months. For each month and each user, we calculate the average \dimension{partisan} score of their activity in that month (i.e.\ the average \dimension{partisan} score of the communities they participated in, weighted by the number of comments they made in each community). We only compute these scores for user/month pairs with at least $10$ comments to minimize noise; results for other choices of threshold are similar and can be found in \figurePolarizationMisc f. \figurePolarizationMisc d shows the Pearson correlation coefficient between the average \dimension{partisan} scores of a user in any pair of months. \figureCohort b shows the proportion of users whose average absolute \dimension{partisan} score increased by $1$ standard deviation for any pair of months. Results for other choices of threshold can be found in \figurePolarizationMisc e.

To measure the effect of individual-level patterns on overall platform polarization, we calculate the average absolute z-score of political activity of new and existing users, and compare these levels of polarization year-over-year. 
A user is considered 'new' at the time of posting a comment if they have no prior activity in political communities 12 months ago or prior.
Let $C_t$ denote the set of all comments in time $t$. Let $E_t$ denote the set of comments made by existing users in time $t$, where a comment $c$ is considered to be made by an existing user if the author of the comment made their first comment in any political community at least 12 months prior to posting $c$.
Let $N_t$ denote the set of comments made by new users in time $t$, i.e.\ all comments not made by existing users ($N_t = C_t \setminus N_t$).  Let $\overline{z}(C)=\frac{1}{|C|}\sum_{c\in C}|z(c)|$ denote the average absolute z-score of comments in set $C$.
The change in average polarization of activity from time $t-1$ to $t$ is equal to $\overline{z}(C_{t})-\overline{z}(C_{t-1})$.
A natural metric to examine the change in average polarization of, for example, new users would be use a similar quantity like $\overline{z}(N_{t})-\overline{z}(N_{t-1})$.
Such a quantity, however, does not itself say anything about platform-level change in polarization, as it does not take into account what proportion of overall activity is made by new or existing users, and whether that proportion itself changed between time periods. In addition, it is an awkward comparison as the new users at time $t$ can be existing users at time $t+1$.
We instead use $\Delta n_{t} = \frac{|N_t|}{|C_t|}(\overline{z}(N_t) - \overline{z}(C_{t-1}))$ to measure the change in \textit{overall} polarization \textit{attributable} to new users.
This represents what the overall change in polarization would have been had the activity of existing users been at the same average level of polarization as that of the platform 12 months prior to when their comment was made. The remaining change is attributable to new user activity and is therefore termed $\Delta n$.
Similarly, $\Delta e_{t} = \frac{|N_t|}{|C_t|}(\overline{z}(E_t) - \overline{z}(C_{t-1}))$ measures the change in polarization attributable to existing users, i.e.\ what the overall change in polarization would have been had the activity of new users been at the same average level of polarization as that of the platform 12 months prior.
This definition also has the desirable property that $\Delta n$ and $\Delta e$ add up to the overall change in polarization, i.e.\ $\Delta n_t + \Delta e_t = \overline{z}(C_{t})-\overline{z}(C_{t-1})$:
\begin{equation}
\begin{split}
\Delta n + \Delta e & = \frac{|N_t|}{|C_t|}(\overline{z}(N_t) - \overline{z}(C_{t-1})) + \frac{|E_t|}{|C_t|}(\overline{z}(E_t) - \overline{z}(C_{t-1}))\\
& = \frac{|N_t|}{|C_t|}\overline{z}(N_t) + \frac{|E_t|}{|C_t|}\overline{z}(E_t) - \frac{|N_t| + |E_t|}{|C_t|}\overline{z}(C_{t-1})\\
& = \overline{z}(C_{t})-\overline{z}(C_{t-1})
\end{split}
\end{equation}
\figureCohort c illustrates the values of $\Delta e$ and $\Delta n$ for each year in our data.

To measure whether polarization patterns differ between left and right wing activity on the platform, we repeat some of the above analyses on two subsets of our data: left-wing activity (including only comments in communities with $z \leq -1$) and right-wing activity (including only comments in communities with $z \geq 1$). We repeat the above change in polarization analysis on the two subsets of data; results can be found in \figureLeftvRight c,d. We repeat the author year-of-first-political-comment analysis on the two subsets of data; results can be found in \figureLeftvRight e,f.

To measure the possible effect of an ``implicit polarization'' process, by which users are influenced by implicitly political subreddits that rank low on the \dimension{partisan-ness} axis but are highly polarized on the \dimension{partisan} axis, we perform an analysis of the relationship between explicitly partisan and implicitly partisan activity.
Examining the 9,453 non-explicitly-political communities, we label communities as 'implicitly political' if they have a \dimension{partisan-ness} score below our cutoff but a \dimension{partisan} score at least 2 standard deviations to the left or right of the global mean in a similar manner to the partisan bins $B$ defined earlier.
We use the sets of explicitly and implicitly partisan communities to examine the relationship between the time users become active in either of them.
Let $m_I(u)$ denote the month that a user $u$ was first active in any implicitly partisan community. Let $m_E(u)$ denote the month that a user $u$ was first active in any explicitly partisan community. 
\figureImplicitPolarization~shows the relationship between $m_I$ and $m_E$ considering both only left-wing activity (left) and right-wing activity (right).
Of users who were first active in an explicitly partisan community at time $m_E$, the proportion of them who were first active in an implicitly partisan community at time $m_I$ is denoted by the colour in cell $(m_E, m_I)$.
The line graphs at the top show the total proportion of users who were active in implicitly partisan communities in a calendar month prior to when they were active in an explicitly partisan community (i.e.\ the proportion of users for whom $m_I < m_E$). 
This corresponds to the proportion of users for which it would be possible for an 'implicit polarization' effect to apply (as it is not possible for an implicit polarization effect to apply if implicitly political activity did not precede explicitly political activity), given that a time granularity of one month is used.

\printbibliography[segment=2,heading=none]
\end{refsegment}

\renewcommand{\figurename}{Appendix Figure}  
\renewcommand{\tablename}{Appendix Table}  
\setcounter{figure}{0}    
\setcounter{table}{0}

\begin{table}[p]
    \centering
    
    \setlength\tabcolsep{3pt}
    {
        \scriptsize
        \sffamily
        \doublespacing
        \begin{tabular}{ c c c c c c c c c }

\multicolumn{6}{c}{\normalsize{\dimension{age}}} &
 \cellcolor{blue!25} teenagers & \cellcolor{blue!25}$\,\to\,$ & \cellcolor{blue!25} RedditForGrownups \\
 
\textsf{youngatheists}& $\,\to\,$ & \textsf{TrueAtheism} & \textsf{teenrelationships}& $\,\to\,$ & \textsf{relationship\_advice} & \textsf{AskMen}& $\,\to\,$ & \textsf{AskMenOver30} \\
\textsf{saplings}& $\,\to\,$ & \textsf{eldertrees} & \textsf{hsxc}& $\,\to\,$ & \textsf{running} & \textsf{trackandfield}& $\,\to\,$ & \textsf{trailrunning} \\
\textsf{TeenMFA}& $\,\to\,$ & \textsf{MaleFashionMarket} & \textsf{bapccanada}& $\,\to\,$ & \textsf{canadacordcutters} & \textsf{RedHotChiliPeppers}& $\,\to\,$ & \textsf{pearljam} \\
    \hline
    
\multicolumn{6}{c}{\normalsize{\dimension{gender}}} &
 \cellcolor{blue!25} AskMen & \cellcolor{blue!25}$\,\to\,$ & \cellcolor{blue!25} AskWomen \\
 
\textsf{TrollYChromosome}& $\,\to\,$ & \textsf{CraftyTrolls} & \textsf{AskMenOver30}& $\,\to\,$ & \textsf{AskWomenOver30} & \textsf{OneY}& $\,\to\,$ & \textsf{women} \\
\textsf{TallMeetTall}& $\,\to\,$ & \textsf{bigboobproblems} & \textsf{daddit}& $\,\to\,$ & \textsf{Mommit} & \textsf{ROTC}& $\,\to\,$ & \textsf{USMilitarySO} \\
\textsf{FierceFlow}& $\,\to\,$ & \textsf{HaircareScience} & \textsf{malelivingspace}& $\,\to\,$ & \textsf{InteriorDesign} & \textsf{predaddit}& $\,\to\,$ & \textsf{BabyBumps} \\
    \hline
    
\multicolumn{6}{c}{\normalsize{\dimension{partisan}}} &
 \cellcolor{blue!25} democrats & \cellcolor{blue!25}$\,\to\,$ & \cellcolor{blue!25} Conservative \\
 
\textsf{GunsAreCool}& $\,\to\,$ & \textsf{progun} & \textsf{OpenChristian}& $\,\to\,$ & \textsf{TrueChristian} & \textsf{GamerGhazi}& $\,\to\,$ & \textsf{KotakuInAction} \\
\textsf{excatholic}& $\,\to\,$ & \textsf{Catholicism} & \textsf{EnoughLibertarianSpam}& $\,\to\,$ & \textsf{ShitRConservativeSays} & \textsf{AskAnAmerican}& $\,\to\,$ & \textsf{askaconservative} \\
\textsf{askhillarysupporters}& $\,\to\,$ & \textsf{AskTrumpSupporters} & \textsf{liberalgunowners}& $\,\to\,$ & \textsf{Firearms} & \textsf{lastweektonight}& $\,\to\,$ & \textsf{CGPGrey} \\
    \hline
    
\multicolumn{6}{c}{\normalsize{\dimension{affluence}}} &
 \cellcolor{blue!25} vagabond & \cellcolor{blue!25}$\,\to\,$ & \cellcolor{blue!25} backpacking \\
 
\textsf{hitchhiking}& $\,\to\,$ & \textsf{hiking} & \textsf{DumpsterDiving}& $\,\to\,$ & \textsf{Frugal} & \textsf{almosthomeless}& $\,\to\,$ & \textsf{personalfinance} \\
\textsf{AskACountry}& $\,\to\,$ & \textsf{travel} & \textsf{KitchenConfidential}& $\,\to\,$ & \textsf{Cooking} & \textsf{Nightshift}& $\,\to\,$ & \textsf{fitbit} \\
\textsf{alaska}& $\,\to\,$ & \textsf{CampingandHiking} & \textsf{fuckolly}& $\,\to\,$ & \textsf{gameofthrones} & \textsf{FolkPunk}& $\,\to\,$ & \textsf{IndieFolk} \\
    \hline
    
\multicolumn{6}{c}{\normalsize{\dimension{age\ B}}} &
 \cellcolor{blue!25} AskMen & \cellcolor{blue!25}$\,\to\,$ & \cellcolor{blue!25} AskMenOver30 \\
 
\textsf{AskWomen}& $\,\to\,$ & \textsf{AskWomenOver30} & \textsf{AskAnAmerican}& $\,\to\,$ & \textsf{RedditForGrownups} & \textsf{androidthemes}& $\,\to\,$ & \textsf{googleplaydeals} \\
\textsf{cringepics}& $\,\to\,$ & \textsf{ghettoglamourshots} & \textsf{windmobile}& $\,\to\,$ & \textsf{canadacordcutters} & \textsf{geegees}& $\,\to\,$ & \textsf{ontario} \\
\textsf{waterpolo}& $\,\to\,$ & \textsf{Yosemite} & \textsf{gatech}& $\,\to\,$ & \textsf{OMSCS} & \textsf{saplings}& $\,\to\,$ & \textsf{eldertrees} \\
    \hline
    
\multicolumn{6}{c}{\normalsize{\dimension{gender\ B}}} &
 \cellcolor{blue!25} daddit & \cellcolor{blue!25}$\,\to\,$ & \cellcolor{blue!25} Mommit \\
 
\textsf{predaddit}& $\,\to\,$ & \textsf{BabyBumps} & \textsf{TallMeetTall}& $\,\to\,$ & \textsf{bigboobproblems} & \textsf{parentsofmultiples}& $\,\to\,$ & \textsf{breastfeeding} \\
\textsf{BeardAdvice}& $\,\to\,$ & \textsf{NoPoo} & \textsf{freemasonry}& $\,\to\,$ & \textsf{pagan} & \textsf{matt}& $\,\to\,$ & \textsf{DrunkOrAKid} \\
\textsf{Leathercraft}& $\,\to\,$ & \textsf{sewing} & \textsf{ketodrunk}& $\,\to\,$ & \textsf{xxketo} & \textsf{techwearclothing}& $\,\to\,$ & \textsf{womensstreetwear} \\
    \hline
    
\multicolumn{6}{c}{\normalsize{\dimension{partisan\ B}}} &
 \cellcolor{blue!25} hillaryclinton & \cellcolor{blue!25}$\,\to\,$ & \cellcolor{blue!25} The\_Donald \\
 
\textsf{GamerGhazi}& $\,\to\,$ & \textsf{KotakuInAction} & \textsf{SandersForPresident}& $\,\to\,$ & \textsf{HillaryForPrison} & \textsf{askhillarysupporters}& $\,\to\,$ & \textsf{AskThe\_Donald} \\
\textsf{BlueMidterm2018}& $\,\to\,$ & \textsf{PoliticalHumor} & \textsf{badwomensanatomy}& $\,\to\,$ & \textsf{ChoosingBeggars} & \textsf{PoliticalVideo}& $\,\to\,$ & \textsf{uncensorednews} \\
\textsf{liberalgunowners}& $\,\to\,$ & \textsf{Firearms} & \textsf{GrassrootsSelect}& $\,\to\,$ & \textsf{DNCleaks} & \textsf{GunsAreCool}& $\,\to\,$ & \textsf{dgu} \\
    \hline
    
\multicolumn{6}{c}{\normalsize{\dimension{sociality}}} &
 \cellcolor{blue!25} nyc & \cellcolor{blue!25}$\,\to\,$ & \cellcolor{blue!25} nycmeetups \\
 
\textsf{law}& $\,\to\,$ & \textsf{LSAT} & \textsf{paris}& $\,\to\,$ & \textsf{travelpartners} & \textsf{sanfrancisco}& $\,\to\,$ & \textsf{SFr4r} \\
\textsf{boston}& $\,\to\,$ & \textsf{bostonhousing} & \textsf{Zappa}& $\,\to\,$ & \textsf{stonerrock} & \textsf{conan}& $\,\to\,$ & \textsf{NewGirl} \\
\textsf{ClashOfClans}& $\,\to\,$ & \textsf{EverWing} & \textsf{answers}& $\,\to\,$ & \textsf{findareddit} & \textsf{xbox360}& $\,\to\,$ & \textsf{XboxOneGamers} \\
    \hline
    
\multicolumn{6}{c}{\normalsize{\dimension{edginess}}} &
 \cellcolor{blue!25} memes & \cellcolor{blue!25}$\,\to\,$ & \cellcolor{blue!25} ImGoingToHellForThis \\
 
\textsf{watchpeoplesurvive}& $\,\to\,$ & \textsf{watchpeopledie} & \textsf{MissingPersons}& $\,\to\,$ & \textsf{MorbidReality} & \textsf{twinpeaks}& $\,\to\,$ & \textsf{TrueDetective} \\
\textsf{pickuplines}& $\,\to\,$ & \textsf{MeanJokes} & \textsf{texts}& $\,\to\,$ & \textsf{FiftyFifty} & \textsf{startrekgifs}& $\,\to\,$ & \textsf{DaystromInstitute} \\
\textsf{subredditoftheday}& $\,\to\,$ & \textsf{SRSsucks} & \textsf{peeling}& $\,\to\,$ & \textsf{Gore} & \textsf{rapbattles}& $\,\to\,$ & \textsf{bestofworldstar} \\
    \hline
    
\multicolumn{6}{c}{\normalsize{\dimension{time}}} &
 \cellcolor{blue!25} PS3 & \cellcolor{blue!25}$\,\to\,$ & \cellcolor{blue!25} PS4 \\
 
\textsf{xbox360}& $\,\to\,$ & \textsf{xboxone} & \textsf{battlefield3}& $\,\to\,$ & \textsf{Battlefield} & \textsf{blackops2}& $\,\to\,$ & \textsf{blackops3} \\
\textsf{deadisland}& $\,\to\,$ & \textsf{dyinglight} & \textsf{ps3bf3}& $\,\to\,$ & \textsf{battlefield\_4} & \textsf{prowrestling}& $\,\to\,$ & \textsf{WWE} \\
\textsf{fo3}& $\,\to\,$ & \textsf{fo4} & \textsf{wii}& $\,\to\,$ & \textsf{wiiu} & \textsf{counter\_strike}& $\,\to\,$ & \textsf{GlobalOffensive} \\
    
        \end{tabular}
    }
    \caption{\protect\textbf{Social dimension seeds.} Community pairs used to calculate social dimensions. The blue highlighted pair is the initial seed provided to the algorithm. The rest of the pairs are algorithmically found as described in Methods.}
    \label{table:appendix_seeds}
\end{table}

\begin{figure}[p]
    \centering
    \includegraphics[width=\textwidth]{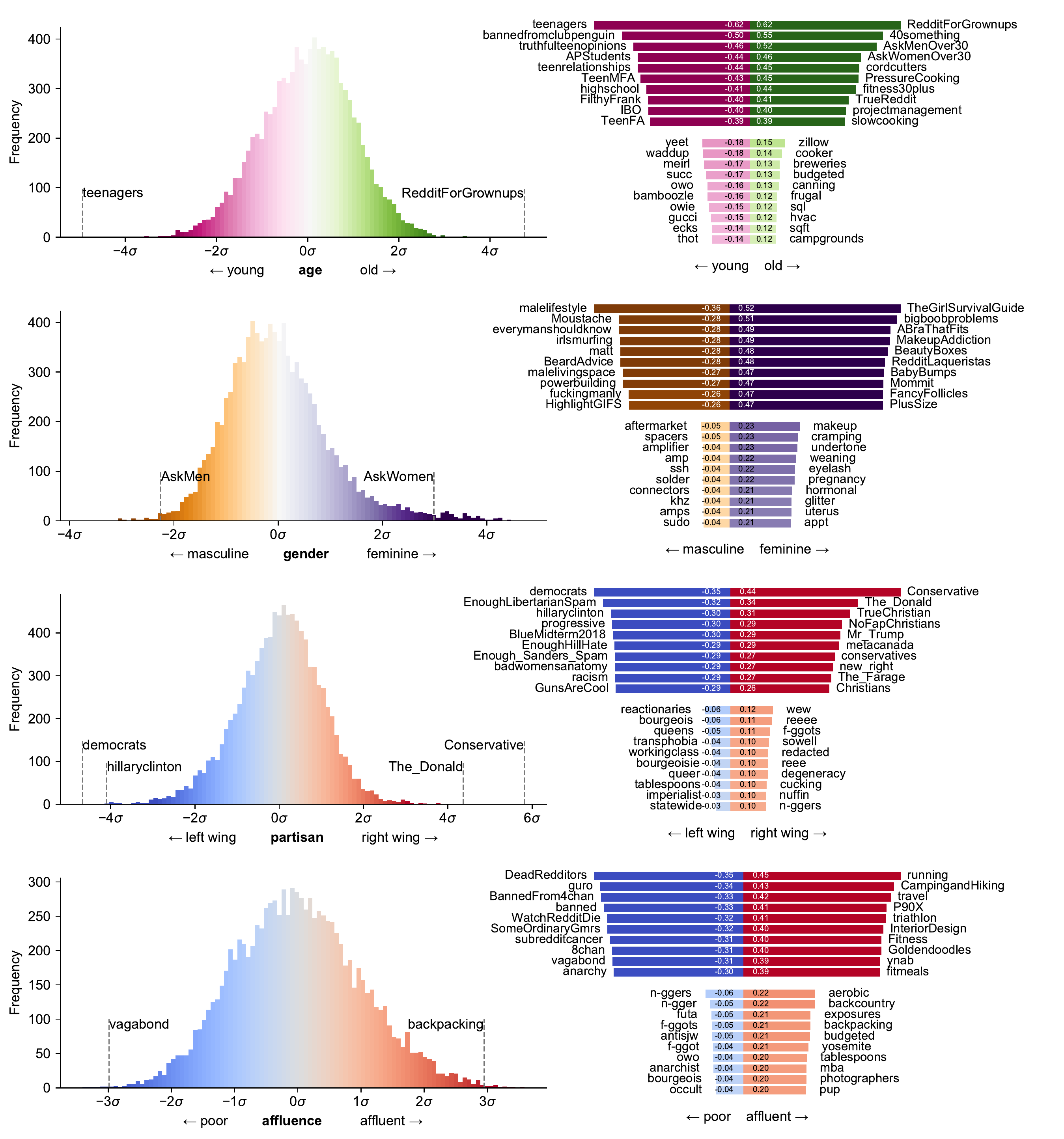}
    \caption{\protect\input{figure-legends/ext1}}
    \label{fig:appendix_dists_1}
\end{figure}

\begin{figure}[p]
    \centering
    \includegraphics[width=\textwidth]{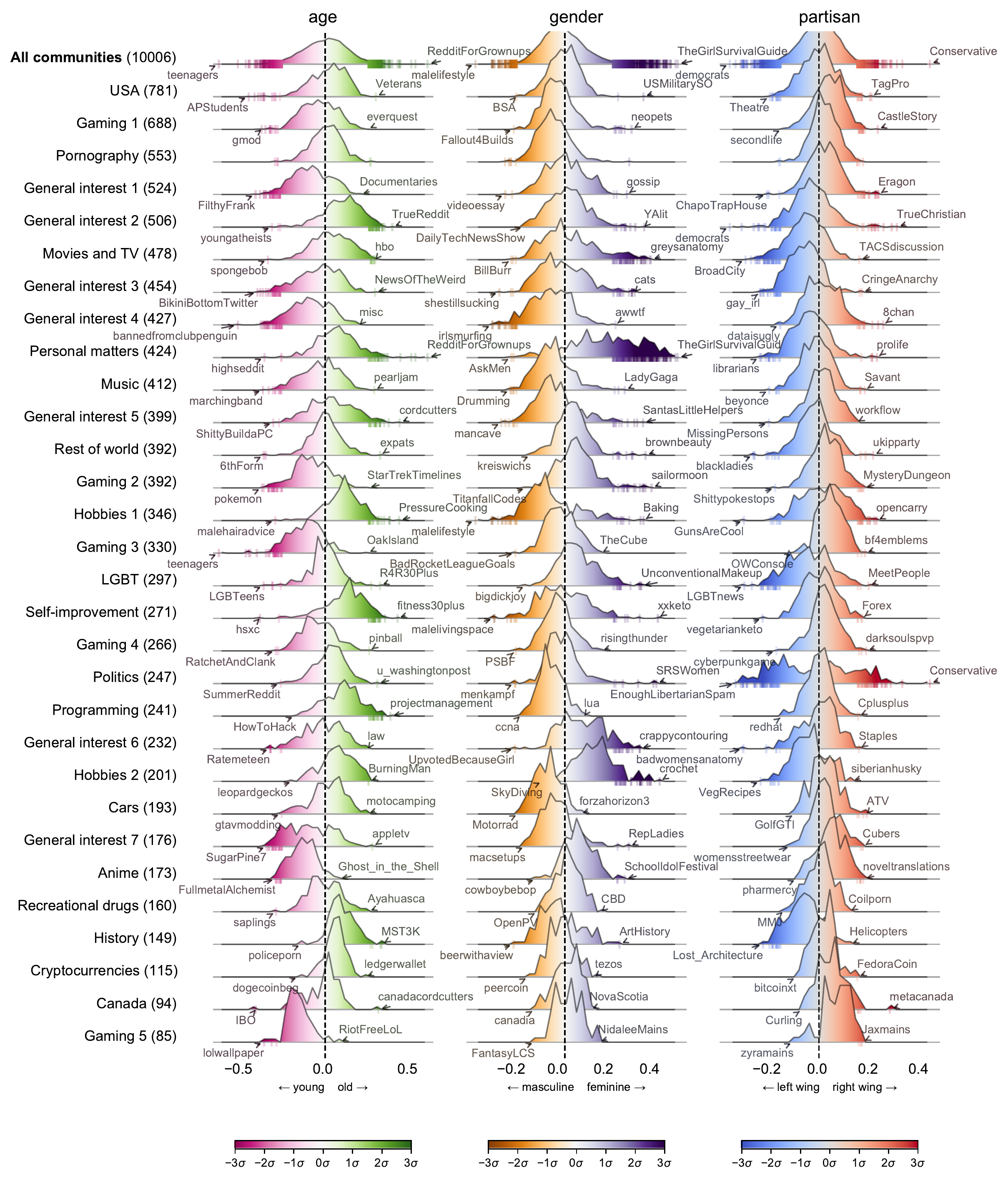}
    \caption{\protect\input{figure-legends/ext2}}
    \label{fig:appendix_densities_0}
\end{figure}

\begin{figure}[p]
    \centering
    \includegraphics[width=\textwidth]{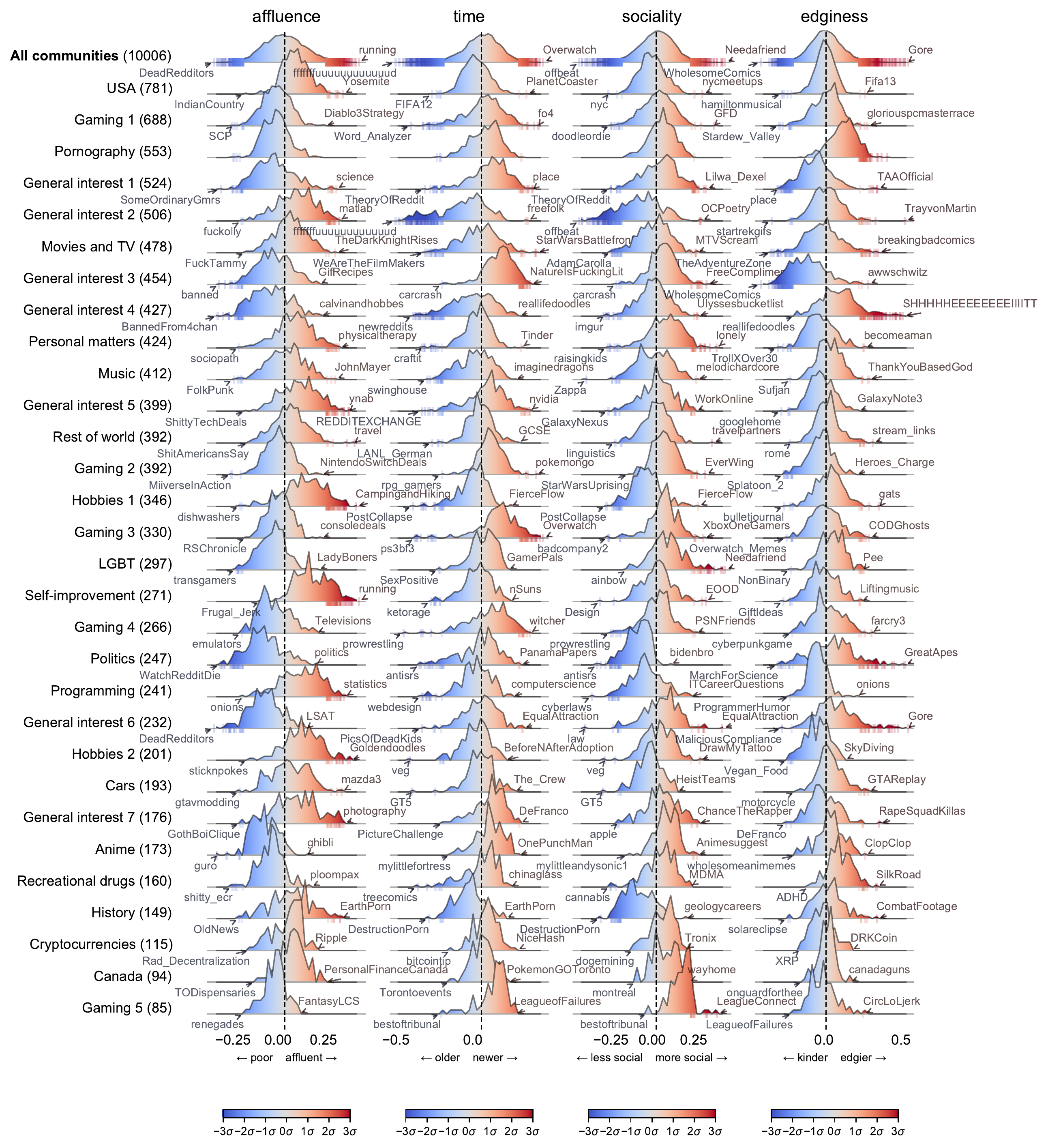}
    \caption{\protect\input{figure-legends/ext3}}
    \label{fig:appendix_densities}
\end{figure}

\begin{figure}[p]
    \centering
    \includegraphics[width=0.5\textwidth]{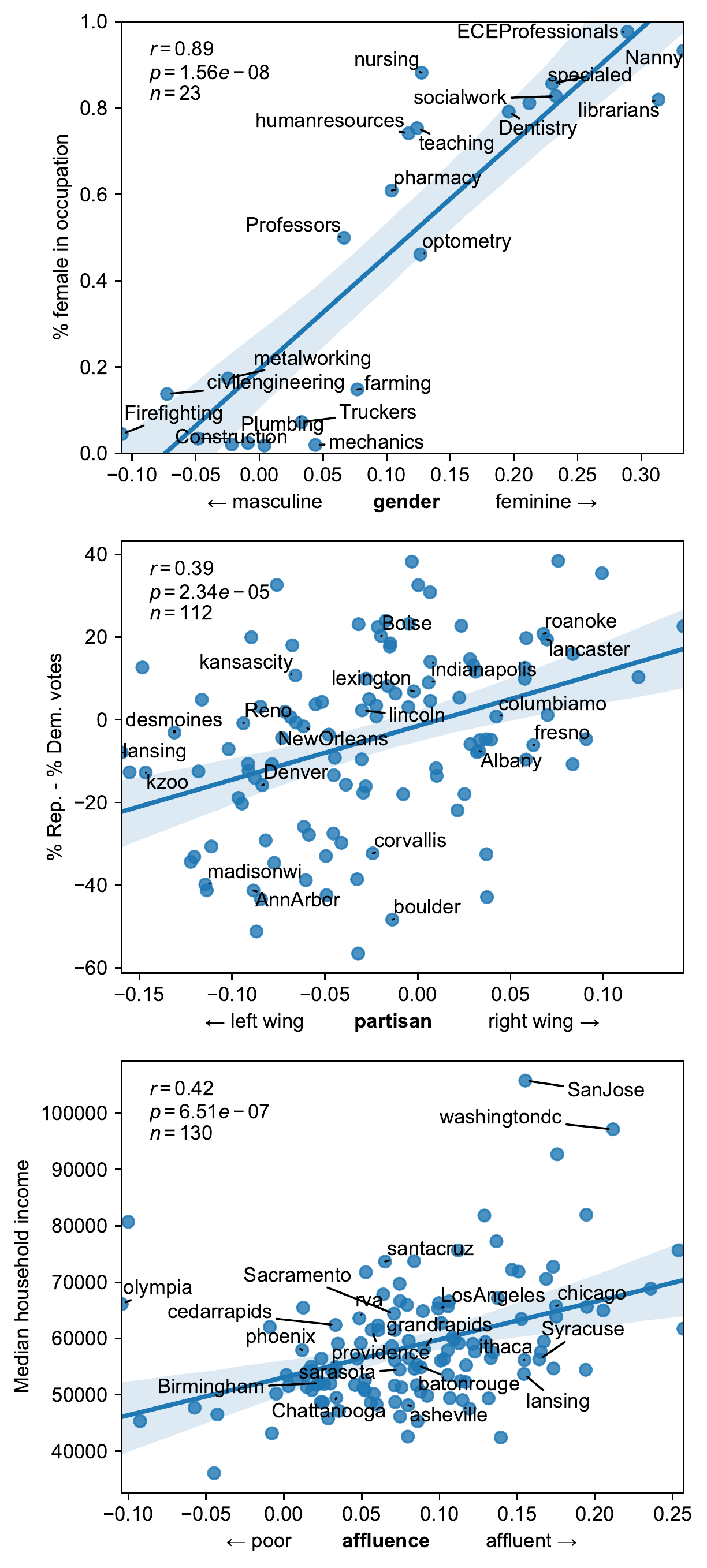}
    \caption{\protect\input{figure-legends/ext4}}
    \label{fig:appendix_validations_2}
\end{figure}

\begin{figure}[p]
    \centering
    \includegraphics[width=0.9\textwidth]{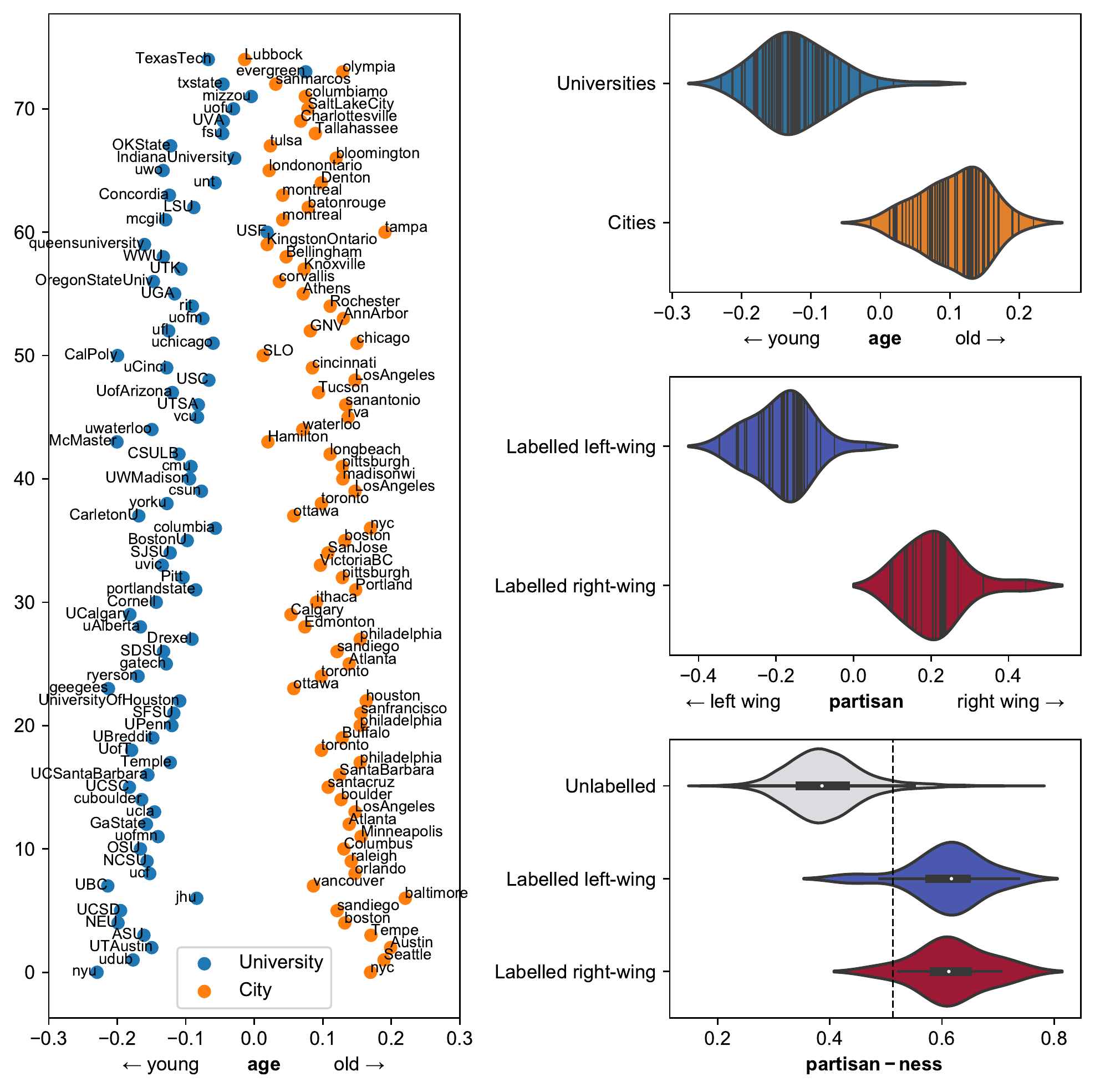}
    \caption{\protect\input{figure-legends/ext5}}
    \label{fig:appendix_validations_1}
\end{figure}

\begin{figure}[p]
    \centering
    \includegraphics[width=\textwidth]{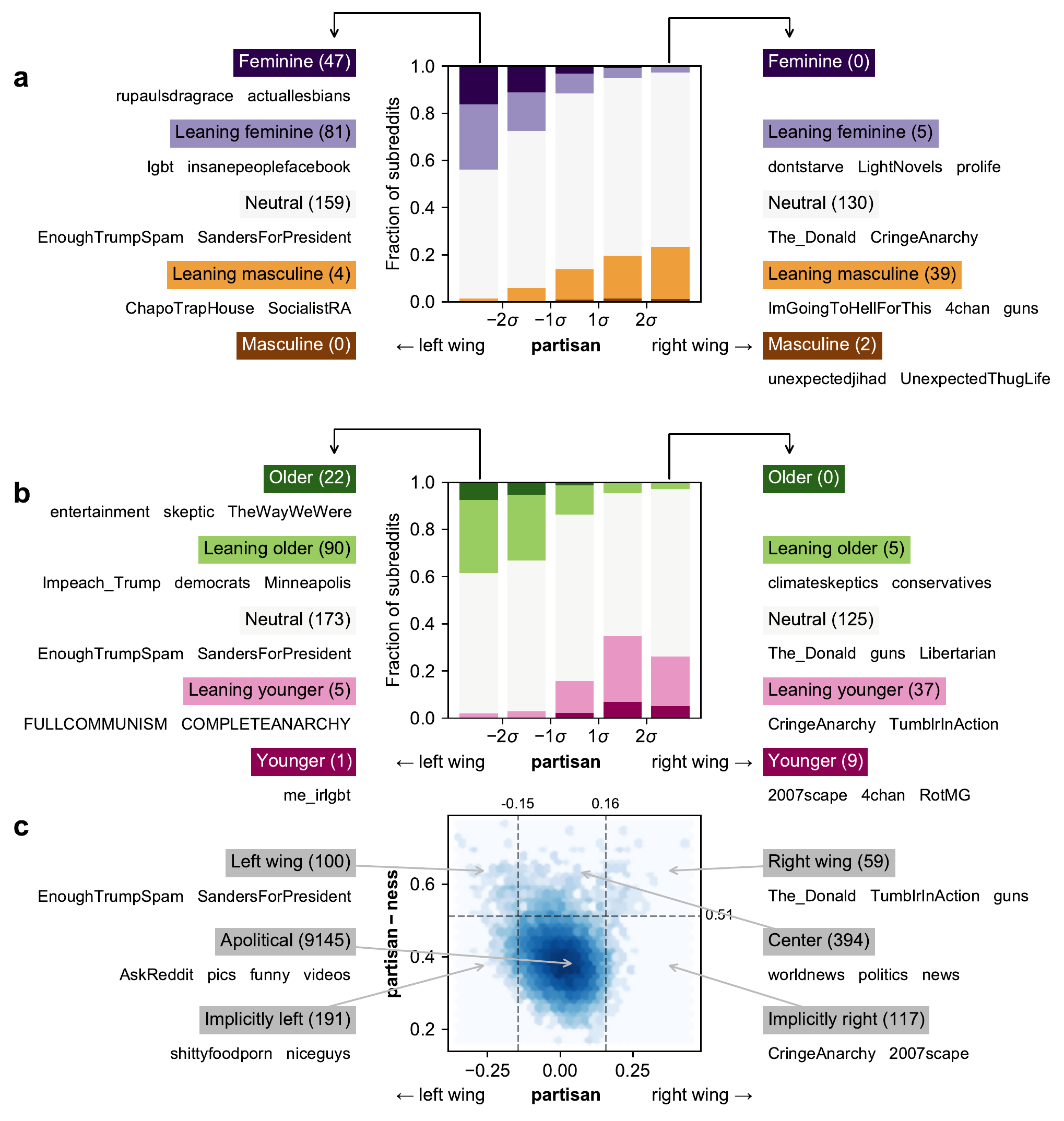}
    \caption{\protect\input{figure-legends/ext6}}
    \label{fig:joint}
\end{figure}

\begin{figure}[h]
    \centering
    \includegraphics[width=\textwidth]{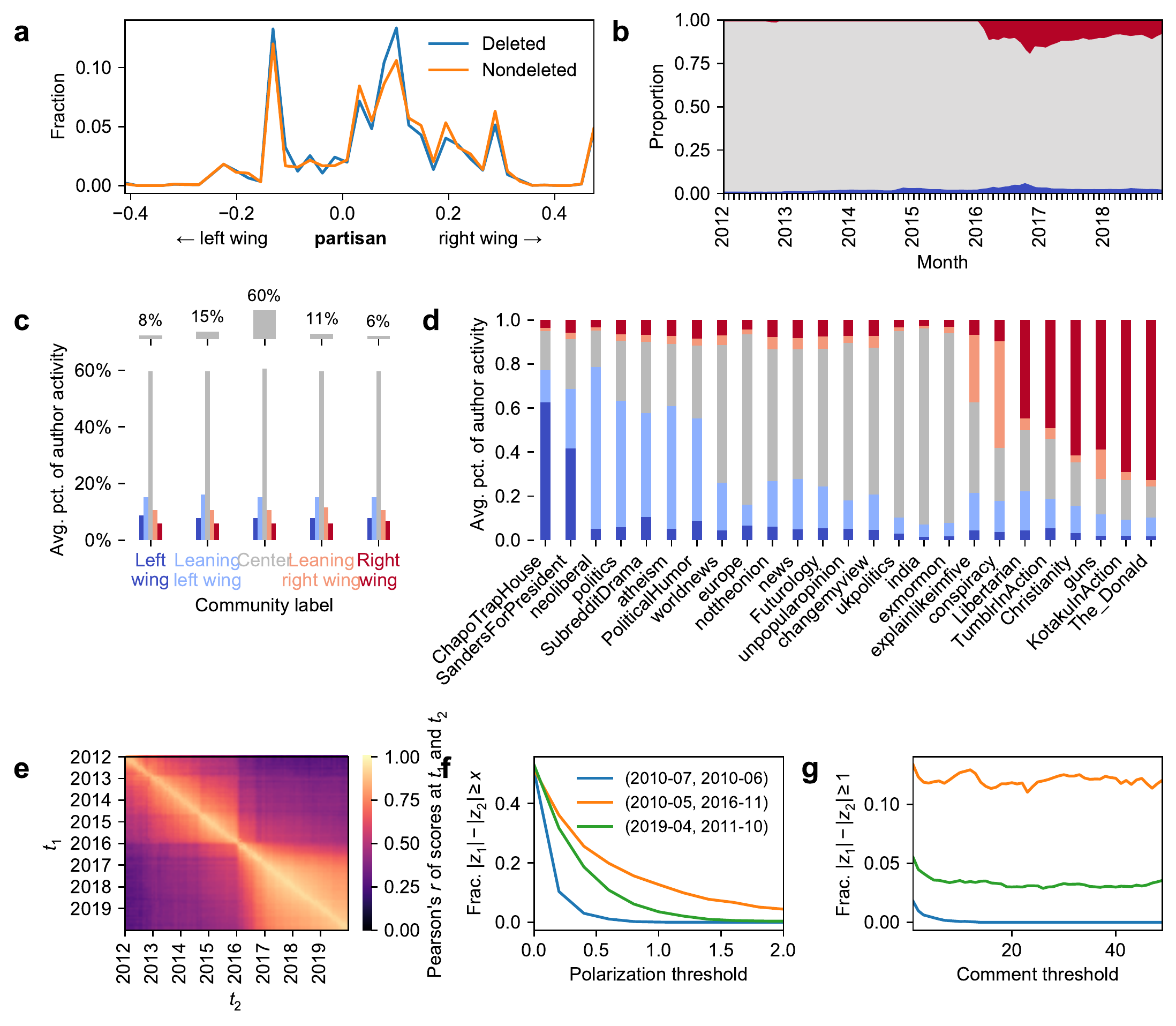}
    \caption{\protect\input{figure-legends/ext7}}
    \label{fig:polarization_additional_plots}
\end{figure}

\begin{figure}[h]
    \centering
    \includegraphics[width=\textwidth]{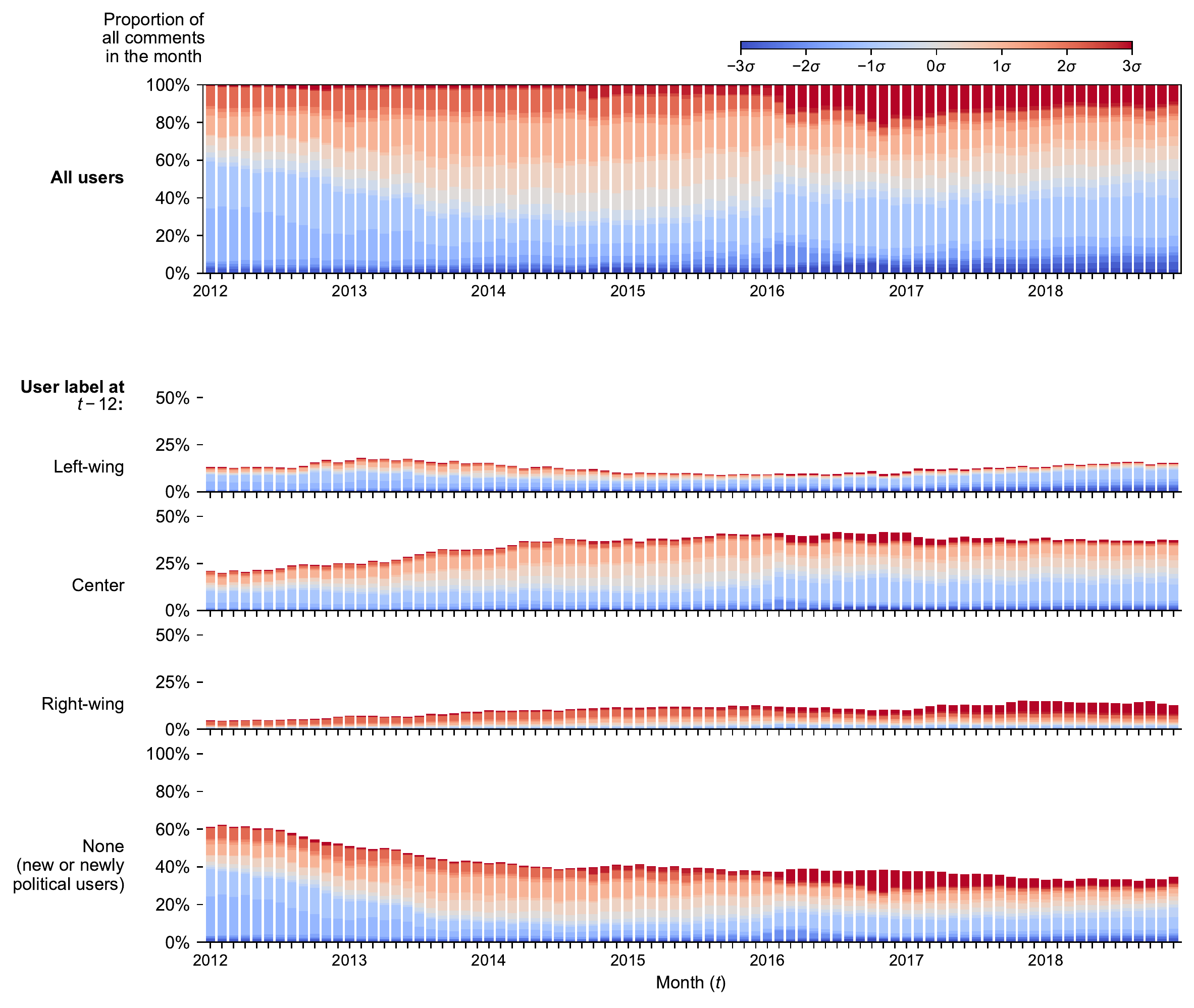}
    \caption{\protect\input{figure-legends/ext8}}
    \label{fig:appendix_orig_activity}
\end{figure}

\begin{figure}[h]
    \centering
    \includegraphics[width=\textwidth]{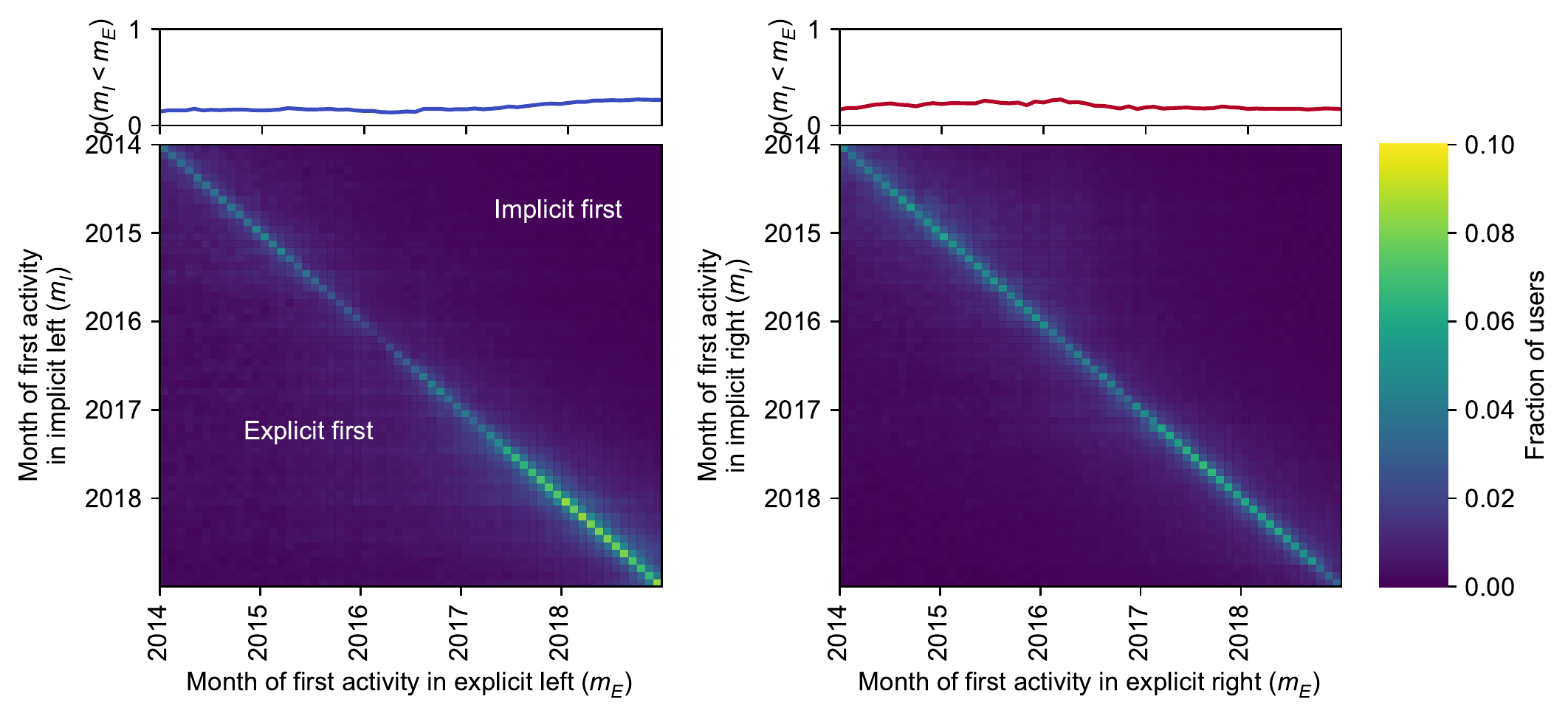}
    \caption{\protect\input{figure-legends/ext9}}
    \label{fig:appendix_implicit_polarization}
\end{figure}

\clearpage

\section*{Supplementary Information}

\section*{Supplementary Table 1: Glossary of referenced communities}
\label{glossary}

Each subreddit is listed along with its description, which is written by the community moderators of the subreddit. Subreddit descriptions were scraped in 2019 and 2020, and as a result, some descriptions are missing for communities which closed or were banned prior to scrape time.

\definecolor{gstd_n3}{rgb}{0.230,0.299,0.754}
\definecolor{gstd_n2}{rgb}{0.436,0.571,0.952}
\definecolor{gstd_n1}{rgb}{0.667,0.779,0.993}
\definecolor{gstd_0}{rgb}{0.867,0.864,0.863}
\definecolor{gstd_1}{rgb}{0.968,0.721,0.612}
\definecolor{gstd_2}{rgb}{0.906,0.455,0.355}
\definecolor{gstd_3}{rgb}{0.706,0.016,0.150}

\definecolor{gstd_age_n3}{rgb}{0.230,0.299,0.754}
\definecolor{gstd_age_n2}{rgb}{0.436,0.571,0.952}
\definecolor{gstd_age_n1}{rgb}{0.667,0.779,0.993}
\definecolor{gstd_age_0}{rgb}{0.867,0.864,0.863}
\definecolor{gstd_age_1}{rgb}{0.968,0.721,0.612}
\definecolor{gstd_age_2}{rgb}{0.906,0.455,0.355}
\definecolor{gstd_age_3}{rgb}{0.706,0.016,0.150}
\definecolor{gstd_gender_n3}{rgb}{0.230,0.299,0.754}
\definecolor{gstd_gender_n2}{rgb}{0.436,0.571,0.952}
\definecolor{gstd_gender_n1}{rgb}{0.667,0.779,0.993}
\definecolor{gstd_gender_0}{rgb}{0.867,0.864,0.863}
\definecolor{gstd_gender_1}{rgb}{0.968,0.721,0.612}
\definecolor{gstd_gender_2}{rgb}{0.906,0.455,0.355}
\definecolor{gstd_gender_3}{rgb}{0.706,0.016,0.150}
\definecolor{gstd_partisan_n3}{rgb}{0.230,0.299,0.754}
\definecolor{gstd_partisan_n2}{rgb}{0.436,0.571,0.952}
\definecolor{gstd_partisan_n1}{rgb}{0.667,0.779,0.993}
\definecolor{gstd_partisan_0}{rgb}{0.867,0.864,0.863}
\definecolor{gstd_partisan_1}{rgb}{0.968,0.721,0.612}
\definecolor{gstd_partisan_2}{rgb}{0.906,0.455,0.355}
\definecolor{gstd_partisan_3}{rgb}{0.706,0.016,0.150}

Each community is labeled with its score on each of our main dimensions, by distance from the mean:

\begin{center}  \scriptsize 
\begin{tabular}{ c | c l | c l | c l }
 $-3\sigma$ & \colorbox{gstd_n3}{A} & Very young &  \colorbox{gstd_n3}{G} & Very masculine & \colorbox{gstd_n3}{P} & Very left-wing \\
 $-2\sigma$ & \colorbox{gstd_n2}{A} & Young &  \colorbox{gstd_n2}{G} & Masculine & \colorbox{gstd_n2}{P} & Left-wing \\
 $-\sigma$ & \colorbox{gstd_n1}{A} & Leaning young &  \colorbox{gstd_n1}{G} & Leaning masculine & \colorbox{gstd_n1}{P} & Leaning left-wing \\
 & \colorbox{gstd_0}{A} & Center &  \colorbox{gstd_0}{G} & Center & \colorbox{gstd_0}{P} & Center \\
 $\sigma$ & \colorbox{gstd_1}{A} & Leaning old &  \colorbox{gstd_1}{G} & Leaning feminine & \colorbox{gstd_1}{P} & Leaning right-wing \\
 $2\sigma$ & \colorbox{gstd_2}{A} & Old &  \colorbox{gstd_2}{G} & Feminine & \colorbox{gstd_2}{P} & Right-wing \\
 $3\sigma$ & \colorbox{gstd_3}{A} & Very old &  \colorbox{gstd_3}{G} & Very feminine & \colorbox{gstd_3}{P} & Very right-wing
\end{tabular}
\end{center}

\setlength\tabcolsep{2pt}

\newcommand{\glossaryheader}[1]{\vspace{0.02cm}\large{\bf #1}}
\newcommand{\glossarytitle}[4]{\scriptsize \textsf{\textbf{#1}} {\colorbox{#2}{A}\colorbox{#3}{G}\colorbox{#4}{P}}}
\newcommand{\glossarydesc}[1]{\scriptsize{#1}}

\begin{singlespace}\begin{flushleft}
\begin{longtabu}[t]{ X X }

\glossaryheader{A} &
\glossarytitle{alaska}{gstd_age_0}{gstd_gender_0}{gstd_partisan_0} \newline \glossarydesc{Subscribed to by 5\% of Alaska's Total Population!}\\
\glossarytitle{almosthomeless}{gstd_age_0}{gstd_gender_1}{gstd_partisan_n1} \newline \glossarydesc{A place to ask for help, advice, or assistance for people who are imminently at-risk of becoming homeless.}&
\glossarytitle{androidthemes}{gstd_age_n1}{gstd_gender_n1}{gstd_partisan_1} \newline \glossarydesc{Showcasing our Android phones, one theme at a time!}\\
\glossarytitle{answers}{gstd_age_1}{gstd_gender_0}{gstd_partisan_0} \newline \glossarydesc{Reference questions answered here.}&
\glossarytitle{askaconservative}{gstd_age_0}{gstd_gender_0}{gstd_partisan_2} \newline \glossarydesc{Ask a Conservative: ask conservatives questions about conservative theory, values, policy and politics. For new conservatives,...}\\
\glossarytitle{AskACountry}{gstd_age_0}{gstd_gender_0}{gstd_partisan_n1} \newline \glossarydesc{Ask countries questions!}&
\glossarytitle{AskAnAmerican}{gstd_age_0}{gstd_gender_0}{gstd_partisan_n1} \newline \glossarydesc{AskAnAmerican: Learn about America, straight from the mouth of Americans}\\
\glossarytitle{askhillarysupporters}{gstd_age_0}{gstd_gender_0}{gstd_partisan_n1} \newline \glossarydesc{Ask supporters of Hillary Clinton for President}&
\glossarytitle{AskMen}{gstd_age_0}{gstd_gender_n2}{gstd_partisan_0} \newline \glossarydesc{r/AskMen is the premier place to ask random strangers for terrible dating advice, but preferably from the male perspective. A...}\\
\glossarytitle{AskMenOver30}{gstd_age_3}{gstd_gender_n1}{gstd_partisan_n1} \newline \glossarydesc{AskMenOver30 is a place for supportive and friendly conversations between over 30 adults....}&
\glossarytitle{AskThe\_Donald}{gstd_age_0}{gstd_gender_0}{gstd_partisan_3} \newline \glossarydesc{A place to discuss Trump, his Administration, and MAGA related topics. Not a safe space.}\\
\glossarytitle{AskTrumpSupporters}{gstd_age_0}{gstd_gender_0}{gstd_partisan_3} \newline \glossarydesc{A Q\&A Subreddit to help improve understanding of the views of Trump Supporters, and the reasons behind those views.}&
\glossarytitle{AskWomen}{gstd_age_0}{gstd_gender_2}{gstd_partisan_n1} \newline \glossarydesc{AskWomen: A subreddit dedicated to asking women questions about their thoughts, lives, and experiences; providing a place where...}\\
\glossarytitle{AskWomenOver30}{gstd_age_3}{gstd_gender_2}{gstd_partisan_n2} \newline \glossarydesc{A place for mature women redditors to discuss questions in a loosely moderated setting.}&\\
\glossaryheader{B} &
\glossarytitle{BabyBumps}{gstd_age_1}{gstd_gender_3}{gstd_partisan_0} \newline \glossarydesc{A place for pregnant redditors, those who have been pregnant, those who wish to be in the future, and anyone who supports them.}\\
\glossarytitle{backpacking}{gstd_age_0}{gstd_gender_n1}{gstd_partisan_0} \newline \glossarydesc{A subreddit for world traveling Backpackers. Often confused with the other type of (wilderness) backpacker in the US, this...}&
\glossarytitle{badwomensanatomy}{gstd_age_0}{gstd_gender_3}{gstd_partisan_n3} \newline \glossarydesc{Women are made of sugar and spice and all things nice.  Except their vaginas which are sqwicky and attract bears.}\\
\glossarytitle{bapccanada}{gstd_age_n1}{gstd_gender_0}{gstd_partisan_0} \newline \glossarydesc{Inspired by a discussion on /r/bapcsalescanada the new /r/bapccanada is the /r/buildapc but for Canadians!}&
\glossarytitle{Battlefield}{gstd_age_n1}{gstd_gender_n1}{gstd_partisan_1} \newline \glossarydesc{/r/battlefield}\\
\glossarytitle{battlefield3}{gstd_age_0}{gstd_gender_n1}{gstd_partisan_1} \newline \glossarydesc{Battlefield3 discussion. Read the rules please. No Memes or blogspam....}&
\glossarytitle{battlefield\_4}{gstd_age_n1}{gstd_gender_n1}{gstd_partisan_1} \newline \glossarydesc{The Official Battlefield 4 Subreddit.}\\
\glossarytitle{BeardAdvice}{gstd_age_0}{gstd_gender_n3}{gstd_partisan_0} \newline \glossarydesc{The definitive source for men who need answers to their bearded questions. Everyone is welcome to join in on the action....}&
\glossarytitle{bestofworldstar}{gstd_age_0}{gstd_gender_n1}{gstd_partisan_1} \newline \glossarydesc{The Best Of World Star Hip-Hop}\\
\glossarytitle{bigboobproblems}{gstd_age_0}{gstd_gender_3}{gstd_partisan_n1} \newline \glossarydesc{Vent in this judgment-free community that encourages discussion in a safe environment.}&
\glossarytitle{blackops2}{gstd_age_n1}{gstd_gender_0}{gstd_partisan_0} \newline \glossarydesc{This Subreddit has moved to /r/CallofDuty!}\\
\glossarytitle{blackops3}{gstd_age_n1}{gstd_gender_0}{gstd_partisan_0} \newline \glossarydesc{Call of Duty: Black Ops III is a military science fiction first-person shooter video game, developed by Treyarch and published by...}&
\glossarytitle{BlueMidterm2018}{gstd_age_0}{gstd_gender_0}{gstd_partisan_n3} \newline \glossarydesc{A subreddit for Democrats to discuss the 2018 midterm elections - primaries, candidates, strategy, news, odds, funding,...}\\
\glossarytitle{boston}{gstd_age_0}{gstd_gender_0}{gstd_partisan_0} \newline \glossarydesc{A reddit for the city of Boston, MA (featuring the cities of Cambridge, Somerville, Malden, Medford, Quincy, Braintree, Newton and...}&
\glossarytitle{bostonhousing}{gstd_age_0}{gstd_gender_0}{gstd_partisan_0} \newline \glossarydesc{r/bostonhousing is a great resource for anyone looking for Boston apartments, rooms for rent in Boston, roommates in Boston,...}\\
\glossarytitle{breastfeeding}{gstd_age_1}{gstd_gender_3}{gstd_partisan_0} \newline \glossarydesc{  This is a community to encourage, support, and educate mothers nursing babies/children through their breastfeeding journey....}&\\
\glossaryheader{C} &
\glossarytitle{CampingandHiking}{gstd_age_1}{gstd_gender_0}{gstd_partisan_0} \newline \glossarydesc{  Hikers who bring Camping gear in their Backpack.     Tips, trip reports, back-country gear reviews, safety and news....}\\
\glossarytitle{canadacordcutters}{gstd_age_2}{gstd_gender_0}{gstd_partisan_0} \newline \glossarydesc{This subreddit is focused on educating Canadians on the legal, reasonably priced options, news and discussion in regards to...}&
\glossarytitle{Catholicism}{gstd_age_0}{gstd_gender_0}{gstd_partisan_2} \newline \glossarydesc{/r/Catholicism is a place to present new developments in the world of Catholicism, discuss theological teachings of the Catholic...}\\
\glossarytitle{CGPGrey}{gstd_age_0}{gstd_gender_0}{gstd_partisan_1} \newline \glossarydesc{Subreddit for CGP Grey stuff.}&
\glossarytitle{ChoosingBeggars}{gstd_age_0}{gstd_gender_0}{gstd_partisan_0} \newline \glossarydesc{NEXT!}\\
\glossarytitle{Christians}{gstd_age_0}{gstd_gender_0}{gstd_partisan_3} \newline \glossarydesc{/r/Christians is a non-denominational community for Christianity that exists firstly for God's glory and secondly for...}&
\glossarytitle{ClashOfClans}{gstd_age_n1}{gstd_gender_0}{gstd_partisan_2} \newline \glossarydesc{Welcome to the subreddit dedicated to the smartphone game Clash of Clans!...}\\
\glossarytitle{conan}{gstd_age_0}{gstd_gender_0}{gstd_partisan_0} \newline \glossarydesc{Sub for Conan talk show on TBS}&
\glossarytitle{Conservative}{gstd_age_0}{gstd_gender_0}{gstd_partisan_3} \newline \glossarydesc{The place for Conservatives on Reddit.}\\
\glossarytitle{conservatives}{gstd_age_1}{gstd_gender_0}{gstd_partisan_3} \newline \glossarydesc{Conservatism (from conservare, "to preserve") is a political and social philosophy that promotes the maintenance of traditional...}&
\glossarytitle{Cooking}{gstd_age_2}{gstd_gender_1}{gstd_partisan_n1} \newline \glossarydesc{/r/Cooking is a place for the cooks of reddit and those who want to learn how to cook. Post anything related to cooking here,...}\\
\glossarytitle{counter\_strike}{gstd_age_0}{gstd_gender_n1}{gstd_partisan_0} \newline \glossarydesc{For the Counter-Strike Gamer. Whether you are a seasoned veteran posting tips, trick and hints or new to the game and need a few...}&
\glossarytitle{CraftyTrolls}{gstd_age_1}{gstd_gender_3}{gstd_partisan_n1} \newline \glossarydesc{Expanding the awesome TrollX and TrollY subreddit universe. Show us your skills! Ask about new ones! Make things!}\\
\glossarytitle{cringepics}{gstd_age_n1}{gstd_gender_0}{gstd_partisan_0} \newline \glossarydesc{An offshoot of /r/cringe, for those images that depict an awkward or embarrassing situation.}&\\
\glossaryheader{D} &
\glossarytitle{daddit}{gstd_age_2}{gstd_gender_n1}{gstd_partisan_0} \newline \glossarydesc{For geek, nerd or neuro-atypical dads.}\\
\glossarytitle{DaystromInstitute}{gstd_age_1}{gstd_gender_0}{gstd_partisan_0} \newline \glossarydesc{A subreddit for in-depth discussion about  Star Trek .}&
\glossarytitle{deadisland}{gstd_age_0}{gstd_gender_0}{gstd_partisan_0} \newline \glossarydesc{First person zombie survival game by Techland....}\\
\glossarytitle{democrats}{gstd_age_1}{gstd_gender_0}{gstd_partisan_n3} \newline \glossarydesc{Offers daily news updates, policy analysis, links, and opportunities to participate in the political process. Feel free to discuss...}&
\glossarytitle{dgu}{gstd_age_1}{gstd_gender_0}{gstd_partisan_1} \newline \glossarydesc{A subreddit dedicated to cataloging incidents in the United States where legally-owned or legally-possessed guns are used to deter...}\\
\glossarytitle{DNCleaks}{gstd_age_1}{gstd_gender_0}{gstd_partisan_0} \newline \glossarydesc{https://wikileaks.org/dnc-emails/  This subreddit was created to post details and significant finds from the DNC leak in a more...}&
\glossarytitle{DrunkOrAKid}{gstd_age_n2}{gstd_gender_0}{gstd_partisan_0} \newline \glossarydesc{This subreddit is for stories of the greatest stupidity.  Inspired by How I Met Your Mother, this subreddit was created for the...}\\
\glossarytitle{DumpsterDiving}{gstd_age_0}{gstd_gender_1}{gstd_partisan_n1} \newline \glossarydesc{Advice, information, and first-hand accounts about finding cool stuff in, or making cool stuff out of, trash.}&
\glossarytitle{dyinglight}{gstd_age_0}{gstd_gender_0}{gstd_partisan_0} \newline \glossarydesc{Dying Light and Dying Light 2 are first person zombie survival games developed by Techland....}\\
\glossaryheader{E} &
\glossarytitle{eldertrees}{gstd_age_1}{gstd_gender_0}{gstd_partisan_n1} \newline \glossarydesc{A friendly haven for ents 18+.}\\
\glossarytitle{Enough\_Sanders\_Spam}{gstd_age_0}{gstd_gender_0}{gstd_partisan_n3} \newline \glossarydesc{Behold! /r/Enough\_Sanders\_Spam, Flame of the Establishment! Forged from the shills of /r/enoughsandersspam.}&
\glossarytitle{EnoughHillHate}{gstd_age_0}{gstd_gender_0}{gstd_partisan_n3} \newline \glossarydesc{\textit{Community banned or deleted prior to publication}}\\
\glossarytitle{EnoughLibertarianSpam}{gstd_age_0}{gstd_gender_0}{gstd_partisan_n3} \newline \glossarydesc{Sick of all the conspiracy theories, racism, anti-Semitism and general douchebaggery of libertarians? You are not alone!   Award...}&
\glossarytitle{EverWing}{gstd_age_0}{gstd_gender_1}{gstd_partisan_1} \newline \glossarydesc{Discussion on anything related to the Facebook Messenger in-app game 'EverWing'.}\\
\glossarytitle{excatholic}{gstd_age_1}{gstd_gender_1}{gstd_partisan_n3} \newline \glossarydesc{This subreddit is for any and all ex-Catholics to talk, educate, discuss and maybe even bitch about their experiences within the...}&\\
\glossaryheader{F} &
\glossarytitle{FierceFlow}{gstd_age_n1}{gstd_gender_n1}{gstd_partisan_0} \newline \glossarydesc{/r/FierceFlow is a subreddit for men with long (er) hair to share tips, progress pictures, anecdotes, or anything else.}\\
\glossarytitle{FiftyFifty}{gstd_age_n2}{gstd_gender_n1}{gstd_partisan_1} \newline \glossarydesc{Risky Clicks the Subreddit}&
\glossarytitle{findareddit}{gstd_age_0}{gstd_gender_0}{gstd_partisan_0} \newline \glossarydesc{Having trouble finding the reddit you need? Post here what you're looking for and someone can suggest a reddit for you!}\\
\glossarytitle{Firearms}{gstd_age_0}{gstd_gender_n1}{gstd_partisan_2} \newline \glossarydesc{A place to discuss firearms and news relating to guns and other small arms. We value the freedom of speech as much as we do the...}&
\glossarytitle{fitbit}{gstd_age_1}{gstd_gender_0}{gstd_partisan_0} \newline \glossarydesc{A forum for discussion of all Fitbit-related products. Come ask questions, encourage/challenge others, and join a community making...}\\
\glossarytitle{fo3}{gstd_age_n1}{gstd_gender_0}{gstd_partisan_0} \newline \glossarydesc{A community for Fallout 3 and everything related....}&
\glossarytitle{fo4}{gstd_age_0}{gstd_gender_0}{gstd_partisan_0} \newline \glossarydesc{The Fallout 4 Subreddit. Talk about quests, gameplay mechanics, perks, story, characters, and more.}\\
\glossarytitle{FolkPunk}{gstd_age_0}{gstd_gender_0}{gstd_partisan_n1} \newline \glossarydesc{A community of punk folks, creating and enjoying folk punk music, and actively standing with Black Lives Matter.}&
\glossarytitle{freemasonry}{gstd_age_1}{gstd_gender_n1}{gstd_partisan_1} \newline \glossarydesc{The main Reddit sub for Masonic Masons and Freemasonry}\\
\glossarytitle{Frugal}{gstd_age_2}{gstd_gender_1}{gstd_partisan_0} \newline \glossarydesc{Frugality is the mental approach we each take when considering our resource allocations. It includes time, money, convenience, and...}&
\glossarytitle{fuckolly}{gstd_age_n1}{gstd_gender_0}{gstd_partisan_0} \newline \glossarydesc{Subreddit dedicated to fucking that piece of shit Olly from HBO's Game of Thrones series. ThanksObama'ed on May 8, but the party...}\\
\glossaryheader{G} &
\glossarytitle{gameofthrones}{gstd_age_0}{gstd_gender_0}{gstd_partisan_0} \newline \glossarydesc{This is a place to enjoy and discuss the HBO series, book series ASOIAF, and GRRM works in general. It is a safe place regardless...}\\
\glossarytitle{GamerGhazi}{gstd_age_0}{gstd_gender_0}{gstd_partisan_n3} \newline \glossarydesc{Diversity and geek culture collide}&
\glossarytitle{gatech}{gstd_age_n1}{gstd_gender_0}{gstd_partisan_0} \newline \glossarydesc{A subreddit for my dear Georgia Tech Yellow Jackets.}\\
\glossarytitle{geegees}{gstd_age_n1}{gstd_gender_0}{gstd_partisan_2} \newline \glossarydesc{University of Ottawa/Université d'Ottawa}&
\glossarytitle{ghettoglamourshots}{gstd_age_1}{gstd_gender_0}{gstd_partisan_0} \newline \glossarydesc{Where Faith in Humanity Comes to Die}\\
\glossarytitle{GlobalOffensive}{gstd_age_n2}{gstd_gender_0}{gstd_partisan_1} \newline \glossarydesc{/r/GlobalOffensive is a home for the Counter-Strike: Global Offensive community and a hub for the discussion and sharing of...}&
\glossarytitle{googleplaydeals}{gstd_age_0}{gstd_gender_0}{gstd_partisan_0} \newline \glossarydesc{The best deals on the Google Play store!}\\
\glossarytitle{Gore}{gstd_age_0}{gstd_gender_0}{gstd_partisan_0} \newline \glossarydesc{\textit{Community banned or deleted prior to publication}}&
\glossarytitle{GrassrootsSelect}{gstd_age_0}{gstd_gender_0}{gstd_partisan_n3} \newline \glossarydesc{  Grassroots Select    has relocated to /r/Political\_Revolution the Reddit branch of The Political Revolution a digital...}\\
\glossarytitle{GunsAreCool}{gstd_age_1}{gstd_gender_0}{gstd_partisan_n3} \newline \glossarydesc{The cost of 'cool'. Mass Shooter Tracker Data.  Mass shootings.  Tracking mass shootings via all guns, firearms, semi-automatics,...}&\\
\glossaryheader{H} &
\glossarytitle{HaircareScience}{gstd_age_0}{gstd_gender_3}{gstd_partisan_0} \newline \glossarydesc{This subreddit aims to provide resources for achieving better hair quality through scientific research in trichology, physiology,...}\\
\glossarytitle{hiking}{gstd_age_1}{gstd_gender_n1}{gstd_partisan_n1} \newline \glossarydesc{The hikers' subreddit.}&
\glossarytitle{hillaryclinton}{gstd_age_0}{gstd_gender_1}{gstd_partisan_n3} \newline \glossarydesc{/r/hillaryclinton is a pro-Hillary Clinton forum to support Hillary Clinton. Join other Hillary Clinton supporters on Reddit! We...}\\
\glossarytitle{HillaryForPrison}{gstd_age_0}{gstd_gender_0}{gstd_partisan_2} \newline \glossarydesc{Hillary Clinton for Prison -- LOCK HER UP!! IT'S WHERE SHE BELONGS. We believe Hillary Clinton should should be in federal prison...}&
\glossarytitle{hitchhiking}{gstd_age_0}{gstd_gender_0}{gstd_partisan_0} \newline \glossarydesc{Good information for the nomadic vagabonds out there. Not just limited to hitchhiking. Trainhopping, destinations, stories, etc....}\\
\glossarytitle{hsxc}{gstd_age_n3}{gstd_gender_n1}{gstd_partisan_1} \newline \glossarydesc{Talk about the season, post race results, and discuss Cross Country in general!}&\\
\glossaryheader{I} &
\glossarytitle{ImGoingToHellForThis}{gstd_age_n1}{gstd_gender_n1}{gstd_partisan_2} \newline \glossarydesc{Looking for a Japanese twink getting railed by two hunky latino men in a hot tub? Trying to figure out the name of the man getting...}\\
\glossarytitle{IndieFolk}{gstd_age_0}{gstd_gender_0}{gstd_partisan_0} \newline \glossarydesc{A Subreddit for Indie \& Folk music. Feel free to post away!}&
\glossarytitle{InteriorDesign}{gstd_age_2}{gstd_gender_1}{gstd_partisan_n1} \newline \glossarydesc{Interior Design is the art and science of understanding people's behavior to create functional spaces within a building.   It is a...}\\
\glossaryheader{K} &
\glossarytitle{ketodrunk}{gstd_age_1}{gstd_gender_0}{gstd_partisan_0} \newline \glossarydesc{A subreddit devoted to the careful craft of the low-carb drunk.  Too many sugary cocktails and carb-laden beer finding their way...}\\
\glossarytitle{KitchenConfidential}{gstd_age_1}{gstd_gender_0}{gstd_partisan_n1} \newline \glossarydesc{Home to the largest community of restaurant and kitchen workers on the internet.}&
\glossarytitle{KotakuInAction}{gstd_age_0}{gstd_gender_0}{gstd_partisan_2} \newline \glossarydesc{KotakuInAction is the main hub for GamerGate on Reddit and welcomes discussion of community, industry and media issues in gaming...}\\
\glossaryheader{L} &
\glossarytitle{lastweektonight}{gstd_age_0}{gstd_gender_0}{gstd_partisan_n3} \newline \glossarydesc{Last Week Tonight with John Oliver is an American late-night talk show airing Sundays on HBO in the United States and HBO Canada,...}\\
\glossarytitle{law}{gstd_age_1}{gstd_gender_0}{gstd_partisan_0} \newline \glossarydesc{A place to discuss developments in the law and the legal profession.}&
\glossarytitle{Leathercraft}{gstd_age_1}{gstd_gender_0}{gstd_partisan_0} \newline \glossarydesc{A subreddit for people interested in working with leather, sharing tips, and tricks.  Learn more about leatherworking and share...}\\
\glossarytitle{liberalgunowners}{gstd_age_1}{gstd_gender_0}{gstd_partisan_n2} \newline \glossarydesc{Gun-ownership through a liberal lens.    This is a place for liberal gun-owners (this means leftist to you "classical liberals")...}&
\glossarytitle{LSAT}{gstd_age_0}{gstd_gender_0}{gstd_partisan_0} \newline \glossarydesc{The best place on Reddit for LSAT advice.   The Law School Admission Test (LSAT) is offered four times per year, and you must...}\\
\glossaryheader{M} &
\glossarytitle{MaleFashionMarket}{gstd_age_0}{gstd_gender_n1}{gstd_partisan_0} \newline \glossarydesc{A place for redditors to sell, buy or trade their previously-owned clothes, shoes and accessories.}\\
\glossarytitle{malelivingspace}{gstd_age_0}{gstd_gender_n2}{gstd_partisan_0} \newline \glossarydesc{MaleLivingSpace is dedicated to places where men can live.   Here you can find posts discussing, showing, improving, and...}&
\glossarytitle{matt}{gstd_age_0}{gstd_gender_n3}{gstd_partisan_0} \newline \glossarydesc{Come ye merry Matts and Matthews!}\\
\glossarytitle{MeanJokes}{gstd_age_n1}{gstd_gender_n1}{gstd_partisan_1} \newline \glossarydesc{A collection of the cruelest, most offensive jokes you can think of.}&
\glossarytitle{memes}{gstd_age_n1}{gstd_gender_0}{gstd_partisan_0} \newline \glossarydesc{Memes!}\\
\glossarytitle{metacanada}{gstd_age_0}{gstd_gender_0}{gstd_partisan_3} \newline \glossarydesc{Canada's only not-retarded subreddit}&
\glossarytitle{MissingPersons}{gstd_age_1}{gstd_gender_1}{gstd_partisan_n2} \newline \glossarydesc{A subreddit for all things related to missing people and their cases.}\\
\glossarytitle{Mommit}{gstd_age_1}{gstd_gender_3}{gstd_partisan_0} \newline \glossarydesc{We are people. Mucking through the ickier parts of child raising. It may not always be pretty, fun and awesome, but we do it....}&
\glossarytitle{MorbidReality}{gstd_age_0}{gstd_gender_1}{gstd_partisan_0} \newline \glossarydesc{Welcome to /r/MorbidReality, a subreddit devoted to the most disturbing content the internet has to offer. Here, we study and...}\\
\glossarytitle{Mr\_Trump}{gstd_age_0}{gstd_gender_n1}{gstd_partisan_3} \newline \glossarydesc{\textit{Community banned or deleted prior to publication}}&\\
\glossaryheader{N} &
\glossarytitle{new\_right}{gstd_age_1}{gstd_gender_0}{gstd_partisan_3} \newline \glossarydesc{  New Right, Alternative Right, Traditionalist, Neoreaction and Dark Enlightenment:   new right news and discussion for...}\\
\glossarytitle{NewGirl}{gstd_age_0}{gstd_gender_1}{gstd_partisan_0} \newline \glossarydesc{A subreddit for fans of the show New Girl. Discussion of, pictures from, and anything else New Girl related.}&
\glossarytitle{Nightshift}{gstd_age_1}{gstd_gender_0}{gstd_partisan_n1} \newline \glossarydesc{For the people who work during the night}\\
\glossarytitle{NoFapChristians}{gstd_age_0}{gstd_gender_n1}{gstd_partisan_3} \newline \glossarydesc{NoFapChristians is a safe place for Christian fapstronauts to discuss the process of abstaining from pornography and masturbation....}&
\glossarytitle{NoPoo}{gstd_age_0}{gstd_gender_1}{gstd_partisan_0} \newline \glossarydesc{"No Shampoo" - A place to discuss natural hair care and alternatives to shampoo. Girls \& Guys welcome!   'No Poo'}\\
\glossarytitle{nyc}{gstd_age_1}{gstd_gender_0}{gstd_partisan_n1} \newline \glossarydesc{r/nyc, the subreddit about new york city}&
\glossarytitle{nycmeetups}{gstd_age_0}{gstd_gender_0}{gstd_partisan_0} \newline \glossarydesc{The subreddit for discussing New York City reddit meetups....}\\
\glossaryheader{O} &
\glossarytitle{OMSCS}{gstd_age_1}{gstd_gender_0}{gstd_partisan_0} \newline \glossarydesc{A place for discussion for people participating in GT's OMS CS}\\
\glossarytitle{OneY}{gstd_age_1}{gstd_gender_n1}{gstd_partisan_0} \newline \glossarydesc{A place to thoughtfully discuss issues that affect men of the world today. Everyone is welcome but intolerance is not.}&
\glossarytitle{ontario}{gstd_age_0}{gstd_gender_0}{gstd_partisan_0} \newline \glossarydesc{A subreddit to discuss all the news and events taking place in the province of Ontario, Canada.}\\
\glossarytitle{OpenChristian}{gstd_age_1}{gstd_gender_1}{gstd_partisan_n2} \newline \glossarydesc{OpenChristian is a community dedicated to Progressive Christianity. This is a space where progressive Christians can support each...}&\\
\glossaryheader{P} &
\glossarytitle{pagan}{gstd_age_1}{gstd_gender_2}{gstd_partisan_0} \newline \glossarydesc{Contemporary Paganism}\\
\glossarytitle{parentsofmultiples}{gstd_age_1}{gstd_gender_0}{gstd_partisan_0} \newline \glossarydesc{A place for parents of twins, triplets, and beyond to discuss the unique challenges of raising and parenting multiples.}&
\glossarytitle{paris}{gstd_age_0}{gstd_gender_0}{gstd_partisan_0} \newline \glossarydesc{La Ville-Lumière. Posts en Anglais et en Français.}\\
\glossarytitle{pearljam}{gstd_age_2}{gstd_gender_0}{gstd_partisan_0} \newline \glossarydesc{A subreddit about all things Pearl Jam. Of course we're talking about the best band from the 1990s featuring Eddie Vedder, Mike...}&
\glossarytitle{peeling}{gstd_age_n1}{gstd_gender_1}{gstd_partisan_0} \newline \glossarydesc{peeling}\\
\glossarytitle{personalfinance}{gstd_age_1}{gstd_gender_0}{gstd_partisan_0} \newline \glossarydesc{Learn about budgeting, saving, getting out of debt, credit, investing, and retirement planning. Join our community, read the PF...}&
\glossarytitle{pickuplines}{gstd_age_n1}{gstd_gender_n1}{gstd_partisan_0} \newline \glossarydesc{A subreddit for all your pick up line needs.  Yes, our icon is a line drawing of a pickup....}\\
\glossarytitle{PoliticalHumor}{gstd_age_0}{gstd_gender_0}{gstd_partisan_n1} \newline \glossarydesc{A subreddit for political humor (particularly US politics), such as political cartoons and satire.}&
\glossarytitle{PoliticalVideo}{gstd_age_0}{gstd_gender_0}{gstd_partisan_0} \newline \glossarydesc{A great place for video content of the political kind. Politics through videos.}\\
\glossarytitle{predaddit}{gstd_age_2}{gstd_gender_n1}{gstd_partisan_0} \newline \glossarydesc{For men about to become fathers}&
\glossarytitle{progressive}{gstd_age_2}{gstd_gender_0}{gstd_partisan_n3} \newline \glossarydesc{A community to share stories related to the growing Modern Political and Social Progressive Movement. The Modern Progressive...}\\
\glossarytitle{progun}{gstd_age_0}{gstd_gender_0}{gstd_partisan_2} \newline \glossarydesc{For pro-gun advocacy!...}&
\glossarytitle{prowrestling}{gstd_age_1}{gstd_gender_0}{gstd_partisan_0} \newline \glossarydesc{  Y  our arena for the enjoyment of the performance art and pseudo-sport aspects of pro wrestling.  Great wrestling from around...}\\
\glossarytitle{PS3}{gstd_age_0}{gstd_gender_0}{gstd_partisan_0} \newline \glossarydesc{  The PlayStation 3 Subreddit (PS3, PlayStation3, Sony PlayStation 3)         The Official FAQ  Your place for:    News   Reviews...}&
\glossarytitle{ps3bf3}{gstd_age_0}{gstd_gender_0}{gstd_partisan_0} \newline \glossarydesc{Battlefield 3 on PS3}\\
\glossarytitle{PS4}{gstd_age_0}{gstd_gender_n1}{gstd_partisan_0} \newline \glossarydesc{The largest PlayStation 4 community on the internet.  Your hub for everything related to PS4 including games, news, reviews,...}&\\
\glossaryheader{R} &
\glossarytitle{racism}{gstd_age_0}{gstd_gender_1}{gstd_partisan_n3} \newline \glossarydesc{Reddit's anti-racism community, a safe(r) space for People of Color and their supporters. All discussions are expected to be from...}\\
\glossarytitle{rapbattles}{gstd_age_0}{gstd_gender_0}{gstd_partisan_0} \newline \glossarydesc{Rap Battles and anything to do with the Battlerap movement around the world.}&
\glossarytitle{RedditForGrownups}{gstd_age_3}{gstd_gender_0}{gstd_partisan_n2} \newline \glossarydesc{This is a community for Redditors that are starting to get that "get off my lawn" feeling whenever they check their front page. So...}\\
\glossarytitle{RedHotChiliPeppers}{gstd_age_n1}{gstd_gender_0}{gstd_partisan_0} \newline \glossarydesc{A community for RHCP fans to share music videos, personal stories, pictures, documentaries, Frusciante solo material, Ataxia, Dot...}&
\glossarytitle{relationship\_advice}{gstd_age_0}{gstd_gender_1}{gstd_partisan_0} \newline \glossarydesc{Need help with your relationship? Whether it's romance, friendship, family, co-workers, or basic human interaction: we're here to...}\\
\glossarytitle{ROTC}{gstd_age_n1}{gstd_gender_n1}{gstd_partisan_1} \newline \glossarydesc{Reserve Officers' Training Corps}&
\glossarytitle{running}{gstd_age_1}{gstd_gender_0}{gstd_partisan_0} \newline \glossarydesc{Runners welcome....}\\
\glossaryheader{S} &
\glossarytitle{SandersForPresident}{gstd_age_0}{gstd_gender_0}{gstd_partisan_n2} \newline \glossarydesc{Midterm Bern \& Bernie 2020! Supporting Senator Bernie Sanders, Our Revolution, National Nurses United, Justice Democrats,...}\\
\glossarytitle{sanfrancisco}{gstd_age_1}{gstd_gender_0}{gstd_partisan_n1} \newline \glossarydesc{Cold summers, thick fog, and beautiful views. Welcome to the subreddit for the gorgeous City by the Bay! San Francisco,...}&
\glossarytitle{saplings}{gstd_age_n2}{gstd_gender_0}{gstd_partisan_0} \newline \glossarydesc{r/saplings: a place to learn about cannabis use and culture}\\
\glossarytitle{sewing}{gstd_age_1}{gstd_gender_3}{gstd_partisan_n1} \newline \glossarydesc{This is a community specifically for sewing including, but not limited to: machine sewing, embroidery, quilting, hand sewing,...}&
\glossarytitle{SFr4r}{gstd_age_0}{gstd_gender_0}{gstd_partisan_n1} \newline \glossarydesc{SFr4r: An R4R fork for the SF Bay Area}\\
\glossarytitle{ShitRConservativeSays}{gstd_age_0}{gstd_gender_0}{gstd_partisan_0} \newline \glossarydesc{Have you ever suspected r/conservative is just liberal satire and most of the moderators are trying to make conservatives look...}&
\glossarytitle{SRSsucks}{gstd_age_0}{gstd_gender_0}{gstd_partisan_1} \newline \glossarydesc{Remember, do not misgender a SRSter, ever. Please refer to them as "It" or "THNG" when "BRD" wont suffice.}\\
\glossarytitle{startrekgifs}{gstd_age_1}{gstd_gender_0}{gstd_partisan_n1} \newline \glossarydesc{A sub to post gifs that are all things Trek}&
\glossarytitle{stonerrock}{gstd_age_1}{gstd_gender_n1}{gstd_partisan_n1} \newline \glossarydesc{Stoner Rock}\\
\glossarytitle{subredditoftheday}{gstd_age_0}{gstd_gender_0}{gstd_partisan_0} \newline \glossarydesc{Subreddit of the Day ... is a celebration of the interesting communities on reddit.com. Once a day we shine a spotlight on the...}&\\
\glossaryheader{T} &
\glossarytitle{TallMeetTall}{gstd_age_0}{gstd_gender_n1}{gstd_partisan_0} \newline \glossarydesc{For tall people who want to meet other tall people}\\
\glossarytitle{techwearclothing}{gstd_age_0}{gstd_gender_n1}{gstd_partisan_0} \newline \glossarydesc{Practical, functional and utilitarian clothing.}&
\glossarytitle{teenagers}{gstd_age_n3}{gstd_gender_0}{gstd_partisan_1} \newline \glossarydesc{r/teenagers is the biggest community forum run by teenagers for teenagers. Our subreddit is primarily for discussions and memes...}\\
\glossarytitle{TeenMFA}{gstd_age_n3}{gstd_gender_0}{gstd_partisan_1} \newline \glossarydesc{\textit{Community banned or deleted prior to publication}}&
\glossarytitle{teenrelationships}{gstd_age_n3}{gstd_gender_0}{gstd_partisan_0} \newline \glossarydesc{A subreddit with the goal is helping teens work through their relationships....}\\
\glossarytitle{texts}{gstd_age_0}{gstd_gender_0}{gstd_partisan_0} \newline \glossarydesc{/r/texts - a subreddit to submit your funny, weird, or random text messages from your mobile or cell phone....}&
\glossarytitle{The\_Donald}{gstd_age_0}{gstd_gender_0}{gstd_partisan_3} \newline \glossarydesc{The\_Donald is a never-ending rally dedicated to the 45th President of the United States, Donald J. Trump.}\\
\glossarytitle{The\_Farage}{gstd_age_0}{gstd_gender_n1}{gstd_partisan_3} \newline \glossarydesc{/r/The\_Farage is the best subreddit of the british god man himself.}&
\glossarytitle{trackandfield}{gstd_age_n2}{gstd_gender_0}{gstd_partisan_1} \newline \glossarydesc{Track and Field}\\
\glossarytitle{trailrunning}{gstd_age_2}{gstd_gender_0}{gstd_partisan_0} \newline \glossarydesc{The fun begins where the road ends....}&
\glossarytitle{travel}{gstd_age_1}{gstd_gender_0}{gstd_partisan_0} \newline \glossarydesc{r/travel is a community about exploring the world. Your pictures, questions, stories, or any good content is welcome.   Clickbait,...}\\
\glossarytitle{travelpartners}{gstd_age_0}{gstd_gender_0}{gstd_partisan_0} \newline \glossarydesc{Meet up with a fellow traveller or travel buddy to visit the world}&
\glossarytitle{TrollYChromosome}{gstd_age_0}{gstd_gender_n1}{gstd_partisan_n1} \newline \glossarydesc{/r/TrollYChromosome is going private to protest against Reddit continuing to provide a platform for racism is hate.    75 Things...}\\
\glossarytitle{TrueAtheism}{gstd_age_1}{gstd_gender_0}{gstd_partisan_n1} \newline \glossarydesc{A subreddit dedicated to insightful posts and thoughtful, balanced discussion about atheism specifically and related topics...}&
\glossarytitle{TrueChristian}{gstd_age_0}{gstd_gender_0}{gstd_partisan_3} \newline \glossarydesc{A subreddit for Christians of all sorts. We exist to be a safe place for discussion between believers on all sides of the fence;...}\\
\glossarytitle{TrueDetective}{gstd_age_1}{gstd_gender_0}{gstd_partisan_0} \newline \glossarydesc{We get the world we deserve.}&
\glossarytitle{twinpeaks}{gstd_age_0}{gstd_gender_0}{gstd_partisan_0} \newline \glossarydesc{A subreddit for fans of David Lynch's and Mark Frost's wonderful and strange television series. We live inside a dream...}\\
\glossaryheader{U} &
\glossarytitle{uncensorednews}{gstd_age_0}{gstd_gender_0}{gstd_partisan_3} \newline \glossarydesc{\textit{Community banned or deleted prior to publication}}\\
\glossarytitle{USMilitarySO}{gstd_age_0}{gstd_gender_3}{gstd_partisan_0} \newline \glossarydesc{Subreddit for sharing advice, support and information for the significant others of current and past members of the United States...}&\\
\glossaryheader{V} &
\glossarytitle{vagabond}{gstd_age_0}{gstd_gender_0}{gstd_partisan_0} \newline \glossarydesc{A digital community created by vagabonds, for vagabonds!   Hitchhikers, Trainhoppers, Rubbertramps, Backpackers, and more!  Feel...}\\
\glossaryheader{W} &
\glossarytitle{watchpeopledie}{gstd_age_0}{gstd_gender_0}{gstd_partisan_0} \newline \glossarydesc{Welcome to watchpeopledie. This community is intended to observe and contemplate the very real reality of death. We are attempting...}\\
\glossarytitle{watchpeoplesurvive}{gstd_age_0}{gstd_gender_n1}{gstd_partisan_n1} \newline \glossarydesc{r/watchpeoplesurvive is like r/watchpeopledie, but dedicated to people surviving near misses, eg:  - Car accidents - Plane crashes...}&
\glossarytitle{waterpolo}{gstd_age_n1}{gstd_gender_n1}{gstd_partisan_1} \newline \glossarydesc{A sports subreddit dedicated to everything water polo related.}\\
\glossarytitle{wii}{gstd_age_0}{gstd_gender_0}{gstd_partisan_0} \newline \glossarydesc{Wii games and scene news}&
\glossarytitle{wiiu}{gstd_age_0}{gstd_gender_0}{gstd_partisan_1} \newline \glossarydesc{Reddit's source for news, pictures, reviews, videos, community insight, \& anything related to Nintendo's 8th-generation console,...}\\
\glossarytitle{windmobile}{gstd_age_0}{gstd_gender_0}{gstd_partisan_0} \newline \glossarydesc{A discussion of WIND Mobile products and services for existing users, those thinking of making the switch, or people just wanting...}&
\glossarytitle{women}{gstd_age_1}{gstd_gender_3}{gstd_partisan_n2} \newline \glossarydesc{A safe, respectful space to discuss the lives and stories of women of all backgrounds, and the current events which affect us....}\\
\glossarytitle{womensstreetwear}{gstd_age_n1}{gstd_gender_2}{gstd_partisan_n1} \newline \glossarydesc{Welcome to Women's Streetwear! Feel free to post any questions, inspirations, pick ups, fit pics, and news regarding women's...}&
\glossarytitle{WWE}{gstd_age_0}{gstd_gender_0}{gstd_partisan_0} \newline \glossarydesc{A subreddit for fans of World Wrestling Entertainment. This includes WCW, ECW, NXT and whatnot.}\\
\glossaryheader{X} &
\glossarytitle{xbox360}{gstd_age_0}{gstd_gender_0}{gstd_partisan_0} \newline \glossarydesc{Everything and anything related to the Xbox 360. News, reviews, previews, rumors, screenshots, videos and more!  Note: We are not...}\\
\glossarytitle{xboxone}{gstd_age_0}{gstd_gender_0}{gstd_partisan_0} \newline \glossarydesc{Everything and anything related to the Xbox One. News, reviews, previews, rumors, screenshots, videos and more!}&
\glossarytitle{XboxOneGamers}{gstd_age_0}{gstd_gender_0}{gstd_partisan_0} \newline \glossarydesc{With many people not upgrading consoles or switching to the other side, friends are hard to find, this subreddit will help make...}\\
\glossarytitle{xxketo}{gstd_age_0}{gstd_gender_3}{gstd_partisan_n1} \newline \glossarydesc{/r/xxketo is a subreddit dedicated to discussing a ketogenic diet from a female-identifying perspective}&\\
\glossaryheader{Y} &
\glossarytitle{Yosemite}{gstd_age_0}{gstd_gender_0}{gstd_partisan_0} \newline \glossarydesc{Yosemite}\\
\glossarytitle{youngatheists}{gstd_age_n2}{gstd_gender_0}{gstd_partisan_0} \newline \glossarydesc{A place for young atheists to share their experiences and questions. We welcome all for a nice slice of skepticism and a cold...}&\\
\glossaryheader{Z} &
\glossarytitle{Zappa}{gstd_age_1}{gstd_gender_0}{gstd_partisan_0} \newline \glossarydesc{All that is Frank Zappa (fl. 1940-1993)}\\

\end{longtabu}
\end{flushleft}
\end{singlespace}

\begin{table}[p]
    \centering
    {\scriptsize \begin{tabular}{ll}
    \toprule\textbf{Community} & \textbf{Occupation} \\
    \midrule
        "Firefighting" & "Firefighters" \\
        \midrule"civilengineering" & "Civil engineers" \\
        \midrule"Construction" & "Construction laborers" \\
        \midrule"metalworking" & \makecell[l]{"Sheet metal workers" \\ "Other metal workers and plastic workers"} \\
        \midrule"Carpentry" & "Carpenters" \\
        \midrule"electricians" & "Electricians" \\
        \midrule"Plumbing" & "Plumbers, pipefitters, and steamfitters" \\
        \midrule"Truckers" & "Driver/sales workers and truck drivers" \\
        \midrule"mechanics" & "Automotive service technicians and mechanics" \\
        \midrule"farming" & "Farmers, ranchers, and other agricultural managers" \\
        \midrule"humanresources" & "Human resources workers" \\
        \midrule"teaching" & \makecell[l]{"Elementary and middle school teachers" \\ "Secondary school teachers"} \\
        \midrule"ECEProfessionals" & "Preschool and kindergarten teachers" \\
        \midrule"nursing" & "Registered nurses" \\
        \midrule"Dentistry" & \makecell[l]{"Dentists" \\ "Dental hygienists" \\ "Dental assistants"} \\
        \midrule"psychotherapy" & \makecell[l]{"Marriage and family therapists" \\ "Therapists, all other"} \\
        \midrule"specialed" & "Special education teachers" \\
        \midrule"socialwork" & \makecell[l]{"Child, family, and school social workers" \\
                    "Mental health and substance abuse social workers" \\
                    "Social workers, all other"} \\
        \midrule"Nanny" & "Childcare workers" \\
        \midrule"optometry" & "Optometrists" \\
        \midrule"pharmacy" & \makecell[l]{"Pharmacists" \\ "Pharmacy aides"} \\
        \midrule"librarians" & "Librarians and media collections specialists" \\
        \midrule "Professors" & "Postsecondary teachers" \\ \bottomrule
    \end{tabular} }
    \caption*{Supplementary Table 2: Reddit community and American Community Survey occupation pairs used for the \dimension{gender} validation. Some subreddits are mapped to multiple occupations, in which case the average gender ratio is used.}
\end{table}

\end{document}